\documentclass[]{aastex631}
\usepackage[makeroom]{cancel}

\begin{document}
\title{Aggregate Cloud Particle Effects in Exoplanet Atmospheres}

\author{Sanaz Vahidinia}
\affiliation{NASA Ames Research Center,
Moffett Field, CA, USA}
\affiliation{NASA Headquarters, Washington, D.C., USA}

\author[0000-0002-6721-3284]{Sarah E. Moran}
\affiliation{Department of Planetary Sciences and Lunar and Planetary Laboratory, University of Arizona, Tuscon, AZ, USA}

\author[0000-0002-5251-2943]{Mark S. Marley}
\affiliation{Department of Planetary Sciences and Lunar and Planetary Laboratory, University of Arizona, Tuscon, AZ, USA}

\author[0000-0003-0553-1436]{Jeff N. Cuzzi}
\affiliation{NASA Ames Research Center, Moffett Field, CA, USA}

\correspondingauthor{Sanaz Vahidinia}
\email{sanaz.vahidinia@nasa.gov}
\correspondingauthor{Sarah E. Moran}
\email{sarahemoran@arizona.edu}

\begin{center}
\vspace{2 mm}
\end{center}



\begin{abstract}
Aerosol opacity has emerged as a critical factor controlling transmission and emission spectra. We provide a simple guideline for the effects of aerosol morphology on opacity and residence time in the atmosphere, as it pertains to transit observations, particularly those with flat spectra due to high altitude aerosols. This framework can be used for understanding complex cloud and haze particle properties before getting into detailed microphysical modeling. We consider high altitude aerosols to be composed of large fluffy particles that can have large residence times in the atmosphere and influence the deposition of stellar flux and/or the emergence of thermal emission in a different way than compact droplet particles as generally modeled to date for extrasolar planetary atmospheres. We demonstrate the important influence of aggregate particle porosity and composition on the extent of the wavelength independent regime. We also consider how such fluffy particles reach such high altitudes and conclude that the most likely scenario is their local production at high altitudes via UV bombardment and subsequent blanketing of the atmosphere, rather than some mechanism of lofting or transport from the lower atmosphere.
 \end{abstract}

\section{Introduction}
Aerosols (clouds and hazes) are found in every major solar system atmosphere and are present in extrasolar planetary atmospheres. The atmospheric thermal profile is affected through the opacity of aerosol particles, which both scatter and absorb light. This scattering and absorption influences the propagation of both incident and emitted radiation. These effects lead to alterations in the shape of resultant emission spectra and the muting of spectral features from gaseous molecules in transmission spectra \citep{seager2010,sing2016}. 

As reviewed in \citet{marley2013}, \citet{marleyrobinson2015}, and \citet{gao2021}, there are a number of different aerosol modeling approaches in use today. Many of these approaches are agnostic as to whether such aerosols are condensate clouds or photochemical hazes. Some models simply define a cloud on an {\it ad hoc} basis (e.g., an arbitrary cloud base and thickness) while others attempt to derive cloud properties on the basis of various physical parameters. In the vast majority of models that treat clouds as non-grey, cloud particles have been assumed to be fully dense, Mie scattering spheres where larger solid particles settle to the base of the cloud, assuming that particles form via condensation \citep[e.g.,][]{AM01}. While the existing modeling approaches have met with a fair amount of success in reproducing spectra and deriving physical parameters for a selection of brown dwarfs \citep[e.g.,][]{cushing2010,Burningham2017,morley2018,miles2023} and extrasolar giant planets \citep[e.g.,][]{Demory2013,Ingraham2014,gao2020,Gao&Powell2021,feinstein2023}, it is more than apparent that systematic differences between models and data remain. Complex  particle morphologies have been considered for decades in the solar system \citep[e.g.,][]{Pollack1980,Toon1980,West1991,Tomasko2008,zhang2013} and over a decade in protoplanetary disks \citep[e.g.,][]{Min2003,Min2005,min2006,kimura2006,Volten2007,Kataoka2014,Cuzzi2014,min2016}, but only recently has particle morphology such as porous aggregates/non-homogeneous particles gained increasing popularity in the exoplanet community \citep{Kopparla2016,Adams2019,Ohno2020,samra2020,Samra2022}, in part due to observations of exoplanets that can be explained by high altitude hazes.

The most convincing evidence of high altitude exoplanetary hazes to date is found in the cases of the transiting planets GJ~1214b and HD~189733b. GJ~1214b is a sub-Neptune orbiting an M star and has a flat transmission spectrum -- the apparent size of the planet as a function of wavelength -- from optical to MIR, across ground-based instruments to the Hubble Space Telescope to JWST. Molecular or atomic absorption signatures as expected from a clear, solar composition atmosphere are not detected \citep{bean2011,DeMooij2013,Kreidberg2014,kempton2023,Gao2023}. In order to flatten spectral features of GJ~1214b,  models with an opaque high altitude cloud or haze layer have been suggested \citep{mr-kempton2012,morley2013,charnay2015,Ohno2018,Lavvas2019,kempton2023}. HD~189733b is a 1.1-Jupiter mass planet orbiting a bright nearby K star and is an excellent target for detailed atmospheric studies. This planet is notable because its transmission spectrum follows a smooth wavelength dependence \citep{sing2011,Pont2013}, suggestive of small particles high in the atmosphere \citep{Gibson2012,Evans2013,Lee2016,ohno&kawashima2020,steinrueck2021}. Similar to GJ~1214b, HD~189733b lacks broad signatures of molecular or atomic absorption at visible wavelengths, although at high spectral resolution, Na and K are detected in the optical \citep{Huitson2012,Pont2013} and CO and H$_2$O are detected in the infrared \citep{brogi2016}.

\begin{figure}[t!]
\centering
{\includegraphics[width=0.99\textwidth]{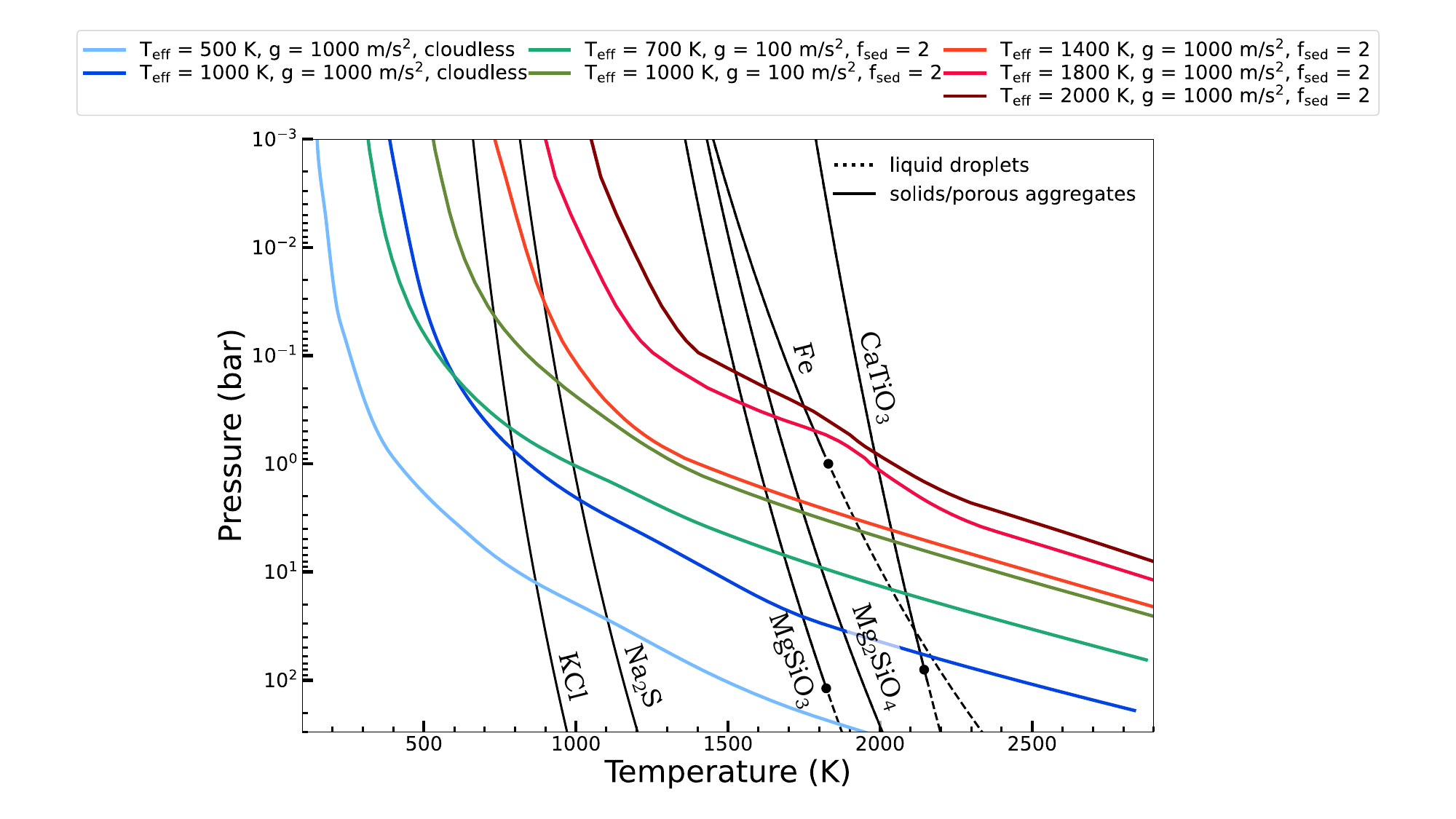}}
\caption{Temperature/pressure diagram showing several substellar atmospheric profiles from \texttt{Sonora Bobcat} \citep{Marley2021} along with the condensation behavior of several representative oxides, salts, silicates, and iron. Line style indicates the phase boundaries \citep[from][]{lodders2009landolt} where the condensate appears as a solid (solid black lines) or a liquid (dotted black lines). Materials condensing as solids will probably form fluffy or porous aggregates, rather than spherical monomers with the density of the pure material.}
\label{fig:PTcomp}
\end{figure}

Since exoplanetary atmospheres can span a wide range of compositions as well as temperature and pressure conditions, and significantly non-cosmic abundances are expected \citep{Moses2013,Fortney2013,Welbanks2019,Bean2021}, a large number of species may form substantial aerosol layers in various phases. Depending on conditions, {these aerosols} in a solar composition atmosphere can include refractory cloud species, such as Fe at high temperatures (hot Jupiters) and Na$_2$S, KCl, ZnS clouds, and complex hydrocarbon hazes at lower temperatures (cooler giants) \citep{Marley1999,marley2013,lodders2009landolt,Zahnle2009,mr-kempton2012,morley2013,Gao&Benneke2018,helling2019,Zhang2020}.  {\bf Figure \ref{fig:PTcomp}} shows condensation curves for a number of important cloud-forming compounds (black curves); the curves are solid where the condensate is predicted to be a solid, and dotted where the condensate is predicted to be a liquid \citep{lodders2009landolt}. The figure also shows temperature-pressure (T-P) profiles (colored lines) for a range of exoplanetary and substellar objects. A cloud base can form at the altitude or pressure where the black and colored curves cross. Particles may settle under gravity, while turbulent mixing can carry particles to higher altitudes, to a degree that depends on their size and density \citep{AM01,marley2013}.  Most cloud-forming species condense as solids, not liquids, over the T-P range of relevance to transmission and emission spectroscopy for exoplanetary atmospheres. 

If particles condense from their vapor phase or are photochemically generated as tiny solids \citep[][and references therein]{helling2006, cable2012}, they coagulate by sticking into porous aggregates \citep[e.g.,][]{Okuzumi2009,Lavvas2011,Adams2019,Ohno2020,yu2021}. We postulate that porous aggregates of solid grains having a wide range of compositions, from refractory condensate clouds to hydrocarbon photochemical hazes, are likely to be \textit{the rule rather than the exception} in the atmospheres of giant exoplanets, as opposed to the currently widespread assumption in most models of spherical monomer particles, such as may be the rule where condensates are liquid. 

\begin{figure}[!ht]
\begin{center}
\includegraphics[width=0.5\textwidth]{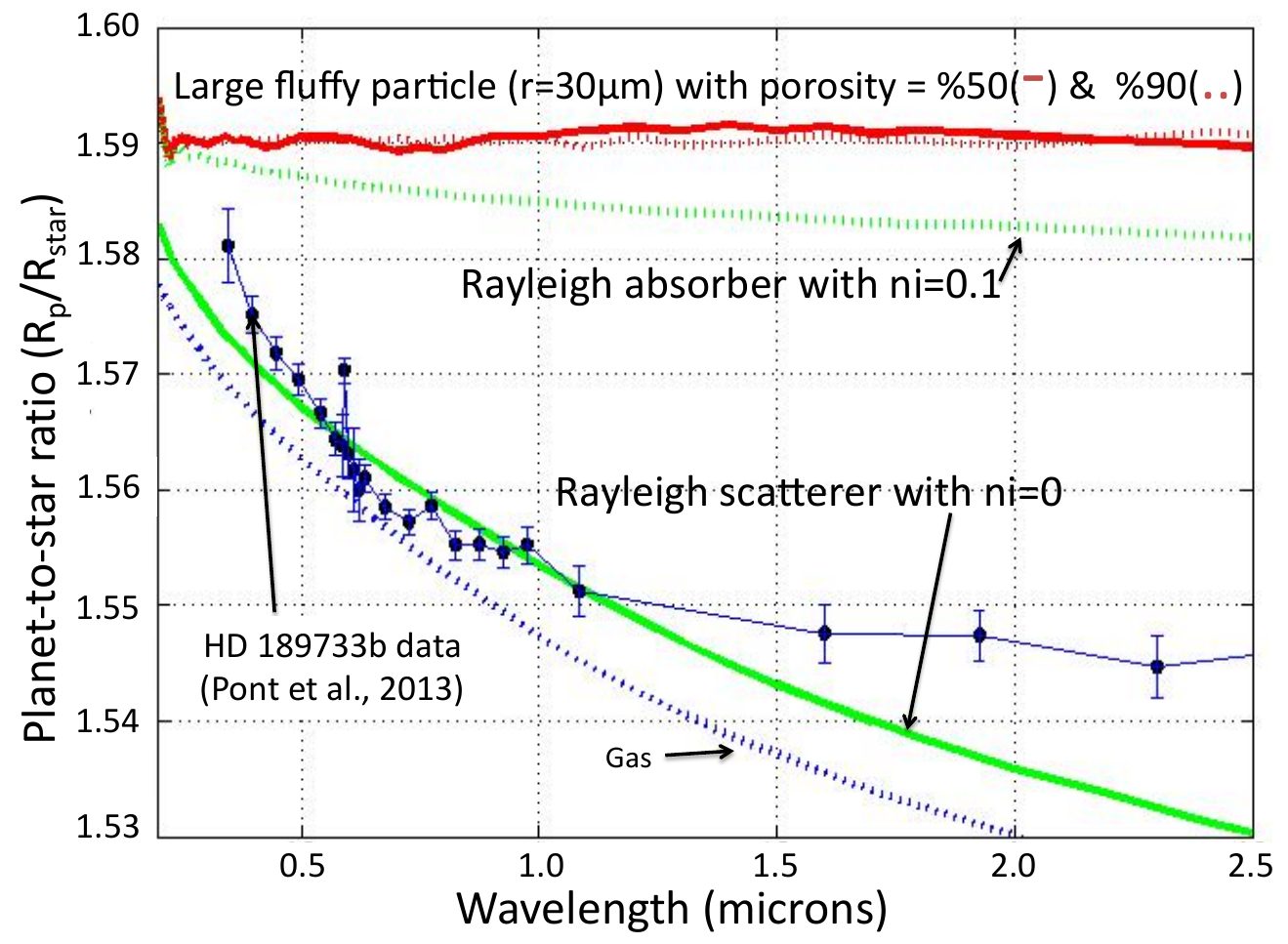}
\caption{Ratio of planet radius $R_\mathrm{p}$ to host star radius $R^*$ vs. wavelength calculated for an atmosphere with extended high altitude aerosols for three different particle types: Red lines show large porous aggregates with a radius of 30 $\mu$m present at high altitudes with 50\% porosity (solid) and 90\% porosity (dotted), producing a flat spectrum (red);  Rayleigh absorbing particles are shown in dotted green; and pure Rayleigh scatterers are shown in solid green, where n$_i$ is the absorbing component of the refractive index of the particle. Data from \citet{Pont2013} (dark blue symbols) and an aerosol-free, gas only atmosphere curve (dotted blue) are also plotted for reference.}
\label{Hazetransit}
\end{center}
\vspace{-0.2in}
\end{figure}

The currently observed flat transit spectra of many exoplanets have been attributed to high altitude aerosols \citep[e.g.,][]{moses2014,Kreidberg2014,Knutson2014}, where the aerosol particles must be larger than the wavelength ($\lambda$) to achieve the flat spectral trend \citep[e.g.,][]{WakefordSing2015,Pinhas2017,KitzmannHeng2018}. The slope of the transmission spectrum depends on the particle size parameter $x$ which is defined as the ratio of particle circumference to the wavelength: $x={2\pi r \over \lambda}$. For particles much smaller than the wavelength -- as is the case for gas phase molecules and the smallest solid particles -- the spectrum follows the Rayleigh curve. As the particles grow, the spectral trend becomes flatter towards the limiting case where the particles are much larger than the wavelength -- where their scattering behavior is wavelength independent. The effects of different scattering regimes and aerosol composition on transit spectra are summarized in \textbf{Figure \ref{Hazetransit}}, where the observed planet-to-star radius ratio ($R_\mathrm{p}\over R_\mathrm{*}$) is calculated for a hot Jupiter with different types of aerosol particles. The planet-to-star ratio for a single aerosol layer can be written as ${R_\mathrm{p}\over R_\mathrm{*}} \sim - H \, \alpha \, ln(\lambda)$ where $H$ is the atmosphere scale height and $\alpha$ defines the scattering regime, which is related to the scattering cross section \citep{Vahidinia2014}. The slope of $R_\mathrm{p}\over R_\mathrm{*}$ ranges from $\alpha=0$ for large fluffy aggregates, to $\alpha = 1$ for Rayleigh absorbing particles, and to $\alpha = 4$ for pure Rayleigh scatterers.

Large particles are needed at high altitudes, but large solid particles settle out quickly. Large porous aggregate particles can resolve this conundrum because they settle much more slowly.
Porosity influences both the radiative properties of these aerosol particles and also their transport and vertical distribution \citep{marley2013}, all of which play determining roles in controlling the observable transmitted spectra of planets.  

 In the next sections we will demonstrate the interplay between particle properties (size, porosity, mass, and composition) that are needed to generate flat spectra at various wavelengths, and what those properties mean in terms of transport and residence times for aerosols in the atmosphere.

\section{Wavelength Independent Regime}

Aerosol opacity ultimately depends upon the radiative properties of the constituent particles. A particle has cross sections to scatter losslessly or to absorb incident radiation, given by $\sigma_{\rm{sca}}$ or $\sigma_{\rm{abs}}$ respectively. These cross sections are defined as $\sigma_{\rm{sca}} = Q_{\rm{sca}} \pi r^2$ and $
\sigma_{\rm{abs}} = Q_{\rm{abs}} \pi r^2$,  where their sum is the extinction cross section $\sigma_\mathrm{ext}$. The scattering and absorption efficiencies $Q_{\rm{sca}}$ and $Q_{\rm{abs}}$ are thus defined, and, from them, the extinction efficiency $Q_\mathrm{ext} = Q_{\rm{sca}} + Q_{\rm{abs}}$, all being functions of the wavelength $\lambda$, through the $\lambda$-dependent real and imaginary refractive indices of the material in question $(n_r, n_i)$; see \citet{DraineLee1984}, \citet{Pollack1994}, or \citet{Cuzzi2014} for typical values. 

To understand the optics of a column of particles, we start with the attenuated light beam $I$ after passing through the column, where $I_\mathrm{0}$ is the incident light: 

\begin{equation}
I=I_\mathrm{0} e^{-\tau} \,\,\,\, \mbox{where the optical depth of the column is defined as  } \tau=n Q_\mathrm{ext} \pi r^{2} H
\label{eq1}
\end{equation}

\begin{figure*}[h!]
\begin{center}
\includegraphics[width=0.8\textwidth]{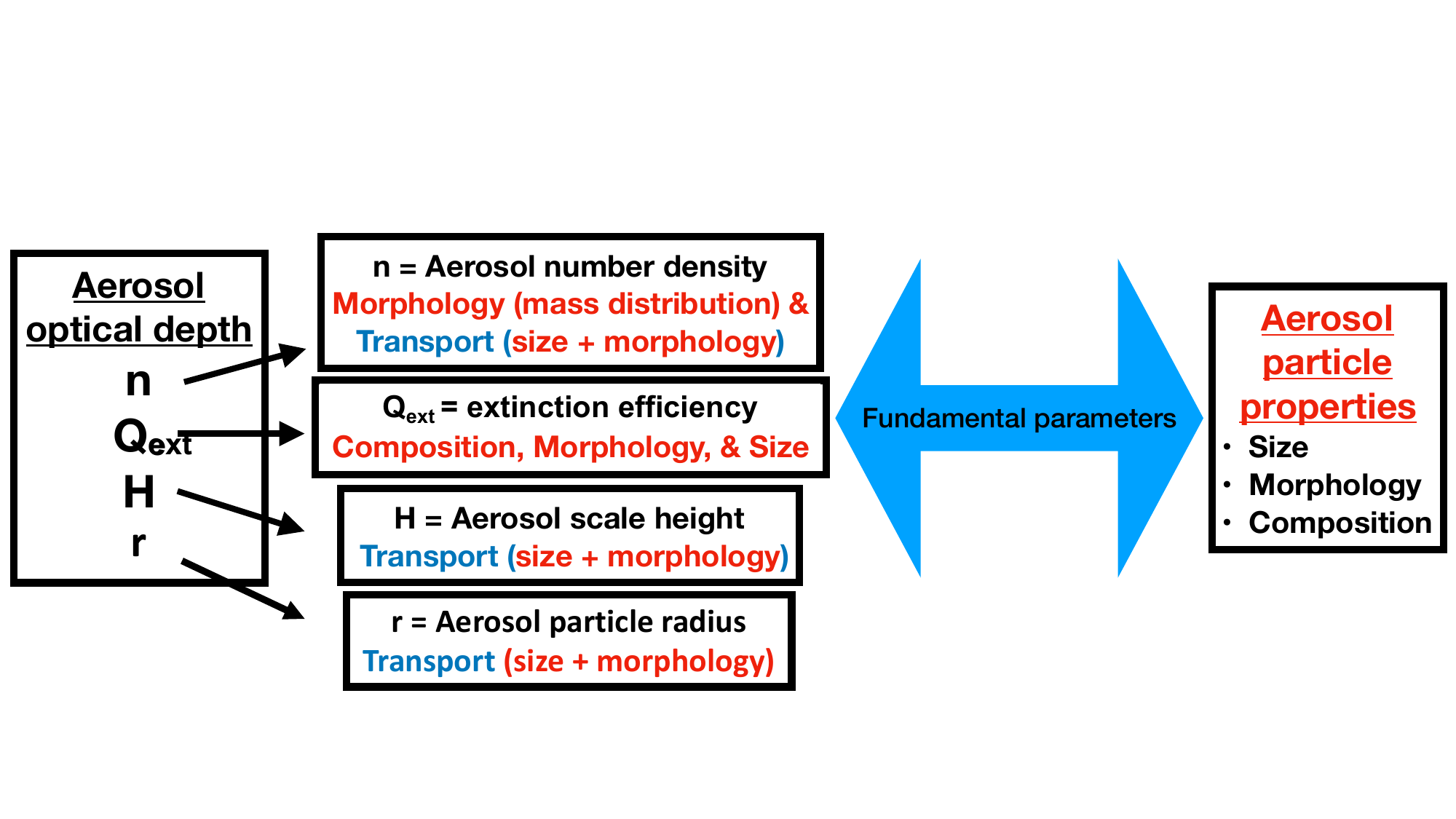}
\end{center}
\caption{Decomposition of optical depth into the fundamental properties of aerosol particles. Particle morphology refers to how mass is distributed within different fractal aggregates shaping the particle cross sectional area and porosity. We will highlight how morphology plays a role in the various components of particle optical depth.}
\label{OD}
\end{figure*}

\noindent The components of optical depth (see {\bf Figure \ref{OD}}) as they pertain to the wavelength independent regime are discussed in detail in the next sections: aerosol extinction efficiency $Q_\mathrm{ext}$, aerosol number density $n$, and aerosol scale height $H$.

The next sections describe the wavelength independent regime of the particle extinction efficiency ($Q_\mathrm{ext}$). The extinction efficiency denotes the amount of energy removed from an incident beam via scattering and absorption and is a major component of the optical depth of a column of atmosphere. The extinction efficiency for homogeneous particles (solid or porous), calculated by using a combination of Mie and Effective Medium theory to mimic porosity, sets the context for establishing the relationship between the extent of wavelength independent regime and particle properties (porosity, composition, and size-to-wavelength ratio, known as the size parameter).

\subsection{Solid Mie particles and the wavelength independent regime} \label{section:solid}
\begin{figure*}[h!]
\begin{center}
\includegraphics[angle=0,width=0.99\textwidth]{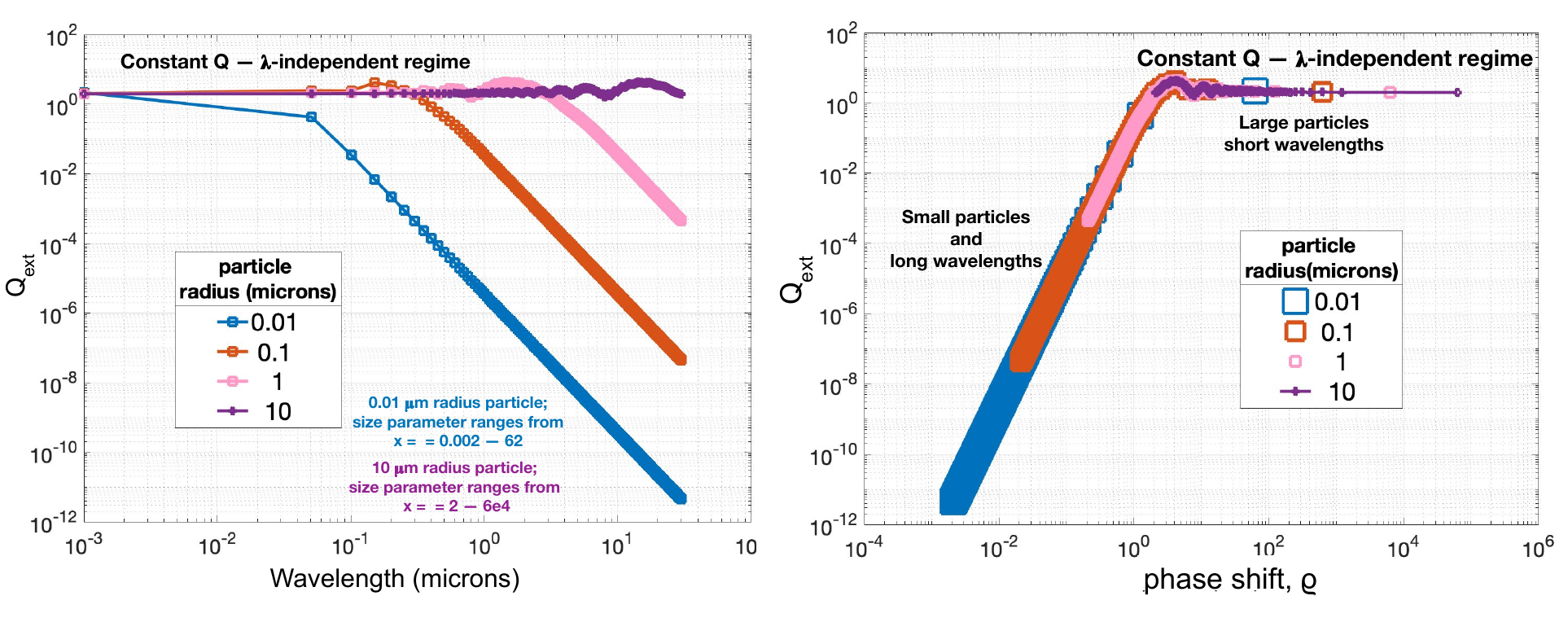}
\label{QexSummary}
\end{center}
\caption{Left: Extinction efficiency ($Q_\mathrm{ext}$) as a function of wavelength $(\lambda)$, shown in different colors for solid particles with radii = 0.01 (blue), 0.1 (red), 1 (pink), and 10 (purple) $\mu$m. When the particles are large compared to the wavelength, the wavelength independent regime is reached. Right: Extinction efficiency ($Q_\mathrm{ext}$) as a function of the optical phase shift ($\varrho$). This is another way of demonstrating constant $Q_\mathrm{ext}$ for different particle sizes where the larger particles maintain constant $Q_\mathrm{ext}$ at longer wavelengths. The phase shift is an important metric that combines size and composition which will be important when we consider particle morphology.}
\label{QE}
\end{figure*}

The wavelength independent regime is simply when the extinction efficiency \textit{$Q_\mathrm{ext}$} is constant as a function of wavelength -- and is often referred to as ``grey clouds'' in the exoplanet literature. Considered physically, the wavelength independent regime depends on the ratio of particle size to wavelength, and the particles' composition and morphology. Before delving into fractal aggregates, important lessons can be learned from homogenous Mie spheres and the range of particle properties that leads to a wavelength independent extinction efficiency ($Q_\mathrm{ext}$). The constant $Q_\mathrm{ext}$ (or wavelength independent) regime is shown in {\bf Figure \ref{QE}} in two different ways: as a function of wavelength $\lambda$, and optical phase shift $\varrho$, which is defined as

\begin{equation}
\varrho =  2 x (n_\mathrm{r} -1) \label{eq:phaseshift}
\end{equation}
where $x$ is the size parameter and $n_\mathrm{r}$ is the real component of the refractive index \citep{hulst1981}. Since the phase shift $\varrho$ is a metric that combines the refractive index and particle size-to-wavelength ratio (i.e., the size parameter), it is used for tracing scattering regimes, such as the transition between Rayleigh regime and geometric optics -- which we call the wavelength independent regime.

Larger particles are able to maintain constant $Q_\mathrm{ext}$ at longer wavelengths as shown in {\bf Figure \ref{QE}} (left panel). Furthermore, there is no unique maximum size after a minimum particle size is reached to be in the wavelength independent regime. As the wavelength increases (e.g., a decade in wavelength space) and observations still show a flat spectrum, the required minimum particle sizes increase (e.g., a decade in size space). 

{\bf 
Figure \ref{QE}} (right panel) shows $Q_\mathrm{ext}$ as a function of the optical phase shift $\varrho$. An important take home message here is that \textit{the optical phase shift needs to be greater than $\sim$several to reach the independent wavelength regime}. There is an interplay between particle size and refractive index which will become important when considering porous particles, which is discussed in the next section (\ref{section:porANDcomp}). The other takeaway is that the observational limit in wavelength dictates the bounds that can be placed on the retrieved particle size.

\subsection{Effects of porosity and composition on the wavelength independent regime} \label{section:porANDcomp}

The most straightforward way of modeling porous particles is to calculate and adopt ``effective'' refractive indices based on their constituent materials and porosity, which are input into a Mie code. If the monomers from which the porous particles are made are smaller than the wavelength in question, they act as independent dipoles immersed in an enveloping medium (where the medium can be another material or vacuum). The porous particle as a whole can then be modeled as having effective refractive indices which depend only on the porosity of the aggregate and the refractive indices (but \textit{not the size}) of the monomers. This is the so-called Effective Medium Theory (EMT); several variants are discussed by \citet{BohrenHuffman1983,ossenkopf1991,Stognienko1995,Voshchinnikov2006}. EMTs can handle either simple one-component, low-density aggregates or physical mixtures of monomers of different composition \citep[e.g.,][]{Helling2008,Cuzzi2014}. {Another common method of accounting for non-spherical or porous particles is the Distribution of Hollow Spheres (DHS) method \citep{Min2003,Min2005}, which is used in several forward model and atmospheric retrieval codes \citep[e.g., \texttt{ARCiS, petitRADTRANS;}][respectively]{Min2020,Molliere2020,Nasedkin2024}. The method of DHS reduces to the same approximation as EMT in the Rayleigh limit, though it is less applicable when particles are large compared to the wavelength of light.}

\begin{figure*}[h!]
\begin{center}
\includegraphics[width=0.99\textwidth]{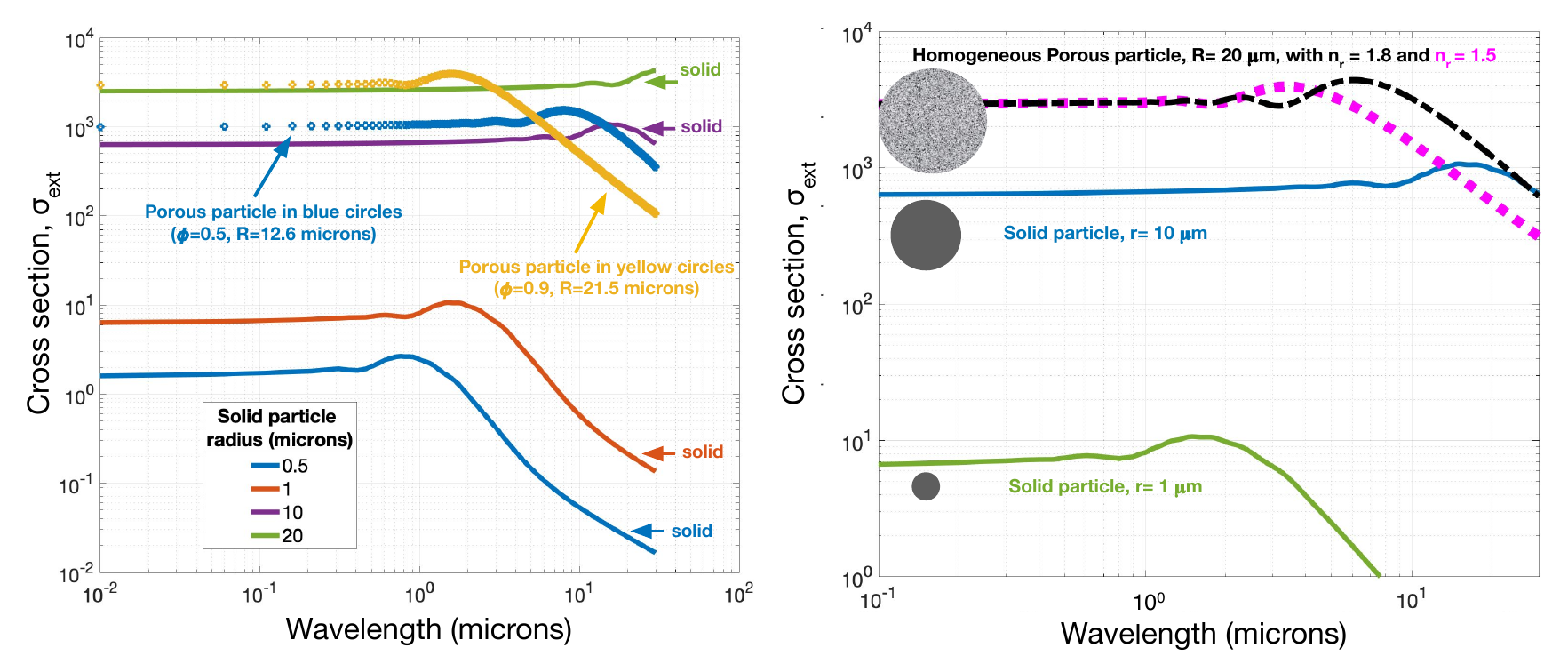}
\end{center}
\caption{Left: scattering cross section for solid particles (solid lines) and homogeneous porous particles (circles) with larger effective radii, $R$, but with the same mass as the $r_s$ = 10 micron solid particle (purple  line). 
The higher porosity causes the cross section to turn over at shorter wavelengths than their solid counterpart of the same mass. The other solid particles are shown for comparison. Right: particle cross sections for a homogeneous porous particle (with porosity $\phi$ = 0.9 and the same mass as that of the solid particle with $r_s$ = 10 microns) at two different refractive indices (dashed lines). The solid particles (with radii of $r_s$ = 1 and $r_s$ = 10 microns, green and blue lines respectively) are shown for reference to demonstrate the turn over from the wavelength independent domain. Note that the left and right panels have different axis limits.}
\label{CE}
\end{figure*}

For this work, we use a simple volume averaged EMT to demonstrate major effects (see Section \ref{section:param_space}). The volume averaged method, or any other variant EMT model in essence lowers the refractive indices for a porous particle compared to its solid component, and these lower refractive indices are used in Mie theory to calculate the scattering properties (see {\bf Figure \ref{CE}}, left panel). {For instance, in the Maxwell Garnett theory \citep{Garnett1904} of EMT, the average refractive index of a porous particle is calculated by assuming that its solid component contains vacuum sites, or that small spherical solid particles are distributed in a vacuum matrix. The amount of vacuum (porosity) is a free parameter in the calculation, where the solid volume fraction is defined as $\mbox{ff}$ and the porosity as $1-\mbox{ff}$.} Therefore, modeling a porous particle using a combination of Mie/EMT methods means calculating the scattering properties of a homogeneous spherical particle with a lower refractive index than that of its solid monomer. The effects of porosity on scattering properties are shown in {\bf Figure \ref{CE}} (left panel) using the combined Mie/volume averaged EMT approach. In these calculations, a solid  particle with radius $r_\mathrm{s}$ is compared to a porous particle of the same mass, which of course has a larger ``effective'' radius, $r_\mathrm{p}$. Thus in the ``short wavelength, large size parameter, large phase shift'' limit, the porous particles have larger extinction cross sections $\sigma_\mathrm{ext}$ than the solid particles, but the same extinction efficiencies $Q_\mathrm{ext}$. The porous particle properties such as mass ($m_\mathrm{p}$), radius ($r_\mathrm{p}$), density ($\rho_\mathrm{p}$), and filling factor ($\mbox{ff}$) are related to those of its solid counterpart ($r_\mathrm{s},\rho_\mathrm{s}$) via the following relationships:  

\begin{equation}
\rho_\mathrm{p}={m_\mathrm{p}\over {4\pi \over 3} r_\mathrm{p}^{3}} = \rho_\mathrm{s}(1-\phi)\mbox{, and } r_\mathrm{p}=r_\mathrm{s}(1-\phi)^{-1/3}
\mbox{and thus the quantity } r_{\rm{p}}\rho_{\rm{p}} = r_{\rm{s}} \rho_{\rm{s}}(1 - \phi)^{2/3} 
\label{same_mass_r}
\end{equation}

where porosity($\phi$)=$1-\mbox{ff}$, $m_\mathrm{p}=m_\mathrm{s}$, and the filling factor ($\mbox{ff}$) is defined as:

 \begin{equation}
\mbox{ff}={\mbox{volume filled}\over \mbox{total volume}}={{4\over3}\pi r_\mathrm{s}^{3} \over {4\over3}\pi r_\mathrm{p}^{3}} = {(r_s / r_p)}^3.
\label{same_mass}
\end{equation}

The product $r_{\rm{p}}$ will appear in boh the radiative transfer and gas dynamics behavior of the particles, tying them together. Higher porosity causes a roll-off from the wavelength independent regime to occur at shorter wavelengths, commensurate with smaller solid particles. In addition to size and porosity, {\it composition} also plays a role in the maximum wavelength up to which the wavelength independent regime extends. Using EMT, we can also express the real {and imaginary refractive indices} of a porous particle ($n_\mathrm{r_p}$,{ $n_\mathrm{i_p}$)} in terms of its solid counterparts ($n_\mathrm{r_s}$,{ $n_\mathrm{i_s}$}):
 \begin{equation}
    (n_\mathrm{r_p} - 1) = (1-\phi)(n_\mathrm{r_s} - 1){ \text{ and } n_\mathrm{i_p} = (1-\phi)(n_\mathrm{i_s})}.
    \label{same_refract}
 \end{equation}
If we apply this to the expression for phase shift in Equation \ref{eq:phaseshift}, we find that
 \begin{equation}
    \varrho_\mathrm{p} = 2x_\mathrm{p}(n_\mathrm{r_s} - 1)(1- \phi),
\label{eq:phaseshift_porous}\end{equation} which more explicitly demonstrates how composition in addition to particle radius affects the particle's extinction. In {\bf Figure \ref{CE}} (right panel), a porous particle with the same mass as the solid particle of radius $r=10$ $\mu$m is shown for two different refractive indices. The porosity makes the particle more transparent and forces the turnover to happen at shorter wavelengths than the solid counterpart. However, as we have shown in Equation \ref{eq:phaseshift_porous}, an increase in the porous particle's refractive index, for example via a different composition or non-homogeneous composition aerosol, pushes the turnover back towards longer wavelengths. A more extended basic treatment of EMT can be found in \citet{Cuzzi2014} (Appendix C) for the interested reader.

\newpage
\subsection{Aerosol scale height, opacity and optical depth }

The particle vertical optical depth through a column of atmosphere ($\tau=n \sigma_\mathrm{ext} H$) is a function of particle number density ($n$), particle cross section ($\sigma_\mathrm{ext}$), and particle scale height ($H$). Here, we have used the cross section \textit{$\sigma_\mathrm{ext}$} rather than the extinction efficiency $Q_\mathrm{ext}$ as in Equation \ref{eq1}. The cross section is simply defined as $\sigma_\mathrm{ext}$ = $Q_\mathrm{ext}$$\pi$$r$$^2$, where \textit{r} is the particle radius. The particle scale height is an important component of optical depth, and porosity can play a major role in the vertical extent of porous particles (with $r_p$) as compared to solid particles (with $r_s$) of the same mass. Following \citet{AM01}, the particle scale height is defined as 

\begin{equation}
H=H_{\rm{g}} {w^{*}\over gt_\mathrm{s}}
\label{particle_scale_height}
\end{equation} 
where $H_{\rm{g}}$ is the gas scale height, $g$ is gravity, $t_\mathrm{s}$ is stopping time which is related to the settling velocity $v_{\rm{f}}$ = $gt_{\rm{s}}$, and $w^{*}$ is the eddy velocity. The particle stopping time is defined as $t_\mathrm{s}= {r \rho \over c \rho_\mathrm{g}}$, where $r$ and $\rho$ are {an aerosol} particle's radius and density, $c$ is the speed of sound in the medium, and $\rho_\mathrm{g}$ is the gas density. Stopping time is discussed in more depth in Section \ref{section:stop_time}. For a porous particle, we have: 

\begin{equation}
H_\mathrm{p} = H_\mathrm{g} {w^{*}\over gt_\mathrm{s_{p}}} = H_\mathrm{g} {w^{*} c \rho_\mathrm{g} \over g r_\mathrm{p} \rho_\mathrm{p}}
\end{equation}

\begin{figure*}[h!]
\begin{center}
\includegraphics[width=0.9\textwidth]{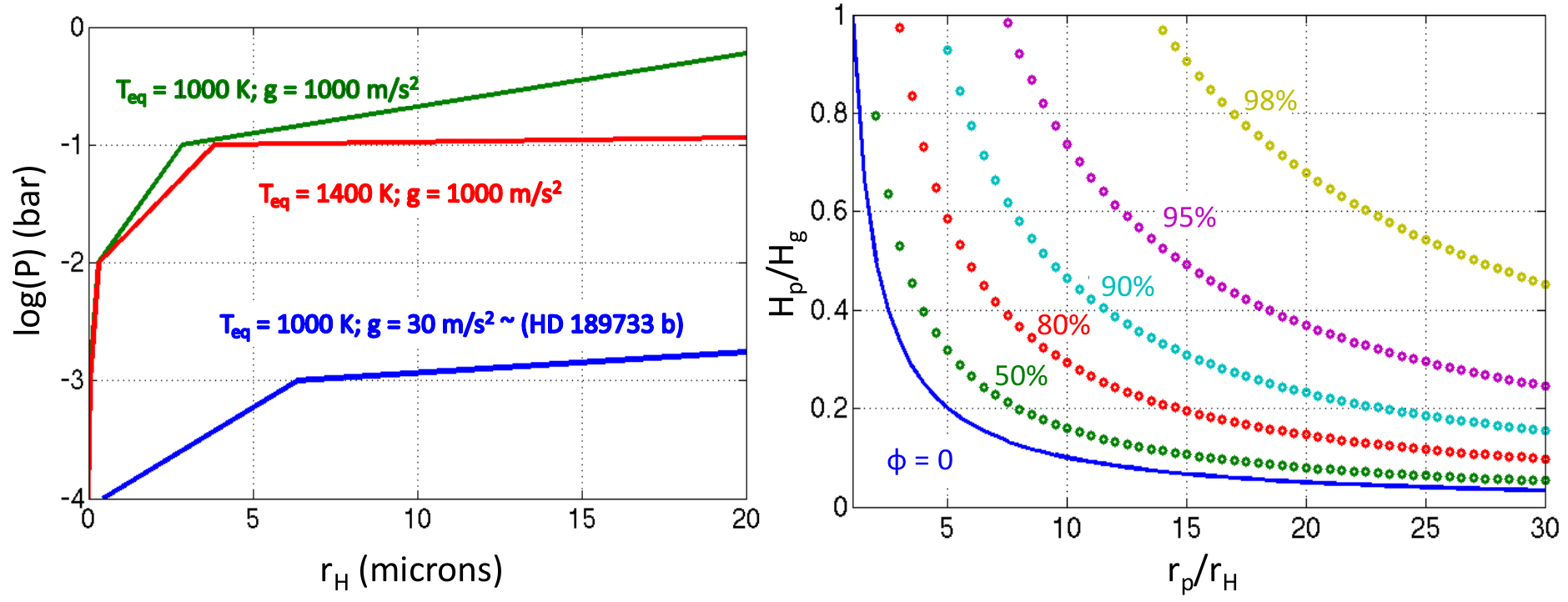}

\caption{Left: particle size scale factor $r_\mathrm{H}$ vs. pressure, calculated by setting $gt_\mathrm{s}=w^{*}$, which gives the maximum particle size that can be suspended as a function of altitude for different planetary gravities and temperatures. Lower gravity planets can have larger cloud particles suspended 
at higher altitudes. Right: ratio of porous particle scale height $H_\mathrm{p}$ to gas scale height $H_\mathrm{g}$ as a function of particle size ratio $r _\mathrm{p}\over r_\mathrm{H}$ for various particle porosities. For example, if a particle size ratio of ${r_\mathrm{p}\over r_\mathrm{H}} \sim 5 $ is needed to flatten a spectrum, then a porosity of approximately 50\% corresponds to a particle scale height comparable to the gas scale height, $H_\mathrm{p}\sim H_\mathrm{g}$.}
\label{porousH}
\end{center}
\end{figure*}

Using the relations in Equation \ref{same_mass_r} for the porous particle density ($\rho_p$) and radius ($r_p$), we can write the porous particle scale height ($H_\mathrm{p}$) relative to a solid particle scale height ($H_\mathrm{s}$) of the same mass:
\begin{equation}
H_\mathrm{p} = H_\mathrm{g} {w^{*} c \rho_\mathrm{g} (1 - \phi)^{-{1 \over 3}} (1 - \phi)^{-{1 \over 3}}  \over g r_\mathrm{s} \rho_\mathrm{s}} =  H_\mathrm{s} (1-\phi)^{-{2 \over 3}}
\label{porous_scaleheight}
\end{equation}
To simplify and illuminate how the particle scale height varies as a function of particle size and porosity, we define a scale factor $r_\mathrm{H}$ to parameterize the porous particle scale height, given by:

\begin{equation}
r_\mathrm{H}={c \rho_\mathrm{g} w^{*} \over g \rho_\mathrm{s}} 
\label{rH}.
\end{equation}
The scale factor $r_\mathrm{H}$ is thus defined as the radius of a {\bf solid} particle having the same scale height as the gas. It is calculated following Equation \ref{particle_scale_height} by setting the settling velocity or terminal velocity ($v_{\rm{f}} = gt_\mathrm{s} = gr_{\rm{s}}\rho_{\rm{s}}/c\rho_{\rm{g}}$) equal to the turbulent velocity ($gt_\mathrm{s}=w^{*}$). That is, the scale factor acts as a tracer for an extended distribution of solid particles maintained at altitude by turbulence against settling. The value of $r_\mathrm{H}$ depends on planetary conditions such as gravity, gas density, temperature (via the density and speed of sound), and eddy diffusivity. If we scale out these factors from Equation \ref{porous_scaleheight}, we can parameterize the particle scale height $H$ as a function of porosity and particle size alone:
\begin{equation}
H_\mathrm{p} = H_\mathrm{g} {r_\mathrm{H}\over r_\mathrm{p}} (1-\phi)^{-1}.
\label{H_por_rh}\end{equation}
Figure \ref{porousH} (left) demonstrates how $r_\mathrm{H}$ varies for three different substellar objects. For example, a planet like a hot Jupiter such as HD~189733b, with lower gravity than a brown dwarf, can sustain larger solid particles higher in its atmosphere. Larger particles (${r_\mathrm{p}\over r_\mathrm{H}} > 1$) can reach high altitudes if their porosity is higher, with an extended scale height that is comparable to the gas scale height (see Figure \ref{porousH}, right).  

Aerosol opacity, $\kappa$ (in units of cm$^{2}$ g$^{-1}$), is defined as the effective particle cross section per unit mass of particles, $m$: 

\begin{equation}
\kappa = {\sigma \over m}
\label{kappa}
\end{equation}

\noindent in the particle or grain. This opacity can be expressed in terms of the total extinction, $\kappa_\mathrm{ext}$, as well as both absorption, $\kappa_\mathrm{abs}$, and scattering, $\kappa_\mathrm{sca}$, with corresponding dependence on the cross section components, $\sigma_\mathrm{ext}$, $\sigma_\mathrm{abs}$, and $\sigma_\mathrm{sca}$. Then the vertical optical depth can be written as a function of extinction opacity and particle extinction cross section:

\begin{equation}
\tau=n \sigma_\mathrm{ext} H= n m \kappa_\mathrm{ext} H.
\label{general_tau}
\end{equation}
If we return to Equation \ref{same_mass_r} and rewrite it in terms of a porous particle's mass, we have:

\begin{equation}
m_\mathrm{p} = {4 \over 3} \pi {r_\mathrm{p}}^3 \rho_\mathrm{s} (1 - \phi) = {4 \over 3} \pi {r_\mathrm{s}}^3 (1 - \phi)^{-1} \rho_\mathrm{s} (1 - \phi) = {4 \over 3} \pi {r_\mathrm{s}}^3 \rho_\mathrm{s} = m_\mathrm{s}.
\end{equation}
Next, along with this expression for the particle mass, we can substitute the porous particle scale height $H_\mathrm{p}$, as expressed in
Equation \ref{porous_scaleheight} as a scaling of the solid particle scale height $H_\mathrm{s}$, into Equation \ref{general_tau} to express the porous particle optical depth as:

\begin{equation}
\tau_\mathrm{p} = n_\mathrm{p} m_\mathrm{p} \kappa_\mathrm{ext_\mathrm{p}} H_\mathrm{p} = n_\mathrm{p} m_\mathrm{s} \kappa_\mathrm{ext_\mathrm{p}} H_\mathrm{s} (1 - \phi)^{-{2 \over 3}} = n_\mathrm{p} \sigma_\mathrm{ext_\mathrm{p}} H_\mathrm{s} (1 - \phi)^{-{2 \over 3}}.
\end{equation}
Exchanging the extinction cross section term for the extinction efficiency, we obtain:

\begin{equation}
\tau_\mathrm{p} =  n_\mathrm{p} Q_\mathrm{ext_\mathrm{p}} \pi r_\mathrm{p}^2 H_\mathrm{s} (1 - \phi)^{-{2 \over 3}}.
\end{equation}
Finally, substituting in our relationship between the solid and porous particle radius from Equation \ref{same_mass_r}, we reach:

\begin{equation}
\tau_\mathrm{p} =  n_\mathrm{p} Q_\mathrm{ext_\mathrm{p}} \pi r_\mathrm{s}^2 (1 - \phi)^{-{2 \over 3}} H_\mathrm{s} (1 - \phi)^{-{2 \over 3}} =  n_\mathrm{p} Q_\mathrm{ext_\mathrm{p}} \pi r_\mathrm{s}^2  H_\mathrm{s} (1 - \phi)^{-{4 \over 3}} .
\end{equation}
Then the ratio of porous to solid optical depth can be written as a ratio of the number densities, extinction efficiencies, and the porosity:

\begin{equation}
{\tau_\mathrm{p}\over \tau_\mathrm{s}}=({n_\mathrm{p} Q_\mathrm{ext_\mathrm{p}}\over n_\mathrm{s} Q_\mathrm{ext_\mathrm{s}}})(1-\phi)^{-{4\over3}}. 
\label{general_tau_ratio}
\end{equation}

To gain insight into how the opacities of solid and porous particles compare, consider the simple case of an opaque particle in the wavelength independent regime. In this case, the particle (solid or porous) is large compared to the wavelength and has an extinction cross section proportional to the geometric cross section ($\sigma_\mathrm{ext}=\pi r^{2}$, where $Q_\mathrm{ext} = {\sigma_\mathrm{ext}\over \pi r^{2}}$), with extinction efficiency $Q_\mathrm{ext} \approx 1$. {Rigorously, $Q_\mathrm{ext} 
 \mbox{~actually} \sim 2$ due to scattering in the short wavelength limit, which can be observed when the angular resolution of the detector is smaller than the deflection of the ray. However, with an extended source -- as is the case with a star -- and an extended screen -- as is the case with a exoplanetary atmosphere -- one will get equivalent light deflected both toward and away from the observer \cite[e.g.,][]{Cuzzi1978,Cuzzi1985}, and thus we can safely treat $Q_\mathrm{ext} \approx 1$.} Therefore, if both porous and solid particles are in the wavelength independent regime and have the same mass, the porous particle has a higher opacity ($\kappa_\mathrm{p}$) compared to that of the solid particle opacity ($\kappa_\mathrm{s}$), as shown in equation \ref{eq_opacity} and discussed in \citet{marley2013}:  

\begin{equation}
\kappa_\mathrm{p}={\pi r_\mathrm{p}^{2}\over ({4\over3} \pi r_\mathrm{p}^{3}\rho_\mathrm{p})}={3\over 4r_\mathrm{s}\rho_\mathrm{s}}(1-\phi)^{-{2\over3}}= \kappa_\mathrm{s} (1-\phi)^{-{2\over3}}.
\label{eq_opacity}
\end{equation}

Here, we have rewritten Equation \ref{kappa} by explicitly expanding out the mass and cross section terms into expressions of the particle radius and then substituted in Equation \ref{same_mass_r}.  Following the same assumption of $Q_\mathrm{ext} \approx 1$, same particle mass criterion that $m_{\rm{s}} = m_{\rm{p}}$, and setting the number density ($n$) to be the same for the solid and porous particles ($n_\mathrm{p} = n_\mathrm{s}$), Equation \ref{general_tau_ratio} can be expressed as Equation \ref{eq_tau}, so that a porous particle has an enhanced optical depth given by: 

\begin{equation}
\tau_\mathrm{p}=\tau_\mathrm{s} (1-\phi)^{-{4\over3}}
\label{eq_tau}
\end{equation}

We can also express the ratio of particle optical depth to gas optical depth if we rewrite the particle number density as $n=\zeta n_\mathrm{g}$, where $\zeta$ is the condensate number abundance and $n_\mathrm{g}$ is the number density of gas molecules with scale height $H$. If we further break down the number abundance as a function of the molecular masses of the particle or condensate ($m_\mathrm{c}$) and gas ($m_\mathrm{g}$), then 

\begin{equation}
    \zeta={m_\mathrm{c} n_\mathrm{molecule}/m_\mathrm{p} \over n_\mathrm{g}} = \zeta_\mathrm{atomic} ({m_\mathrm{c}\over m_\mathrm{p}})
\label{zeta_eq}\end{equation}
 The atomic number density $\zeta_\mathrm{atomic}={n_\mathrm{molecule}\over n_\mathrm{g}}$ is defined as the ratio of the number density of the condensate molecules to that of gas. \textbf{Please Note:} the particle number density ($n$) and particle mass ($m_\mathrm{p}$) are \textbf{not} the same as the {\bf molecular} 
 number density ($n_\mathrm{molecule}$)  and the {\bf molecular} mass ($m_\mathrm{c}$)  of the condensate that makes up the particle. If we substitute $\zeta_\mathrm{atomic}$ and the molecular masses of the particle ($m_\mathrm{p}$) and gas ($m_\mathrm{g}$) from Equation \ref{zeta_eq} into Equation \ref{general_tau} ($\tau= n m \kappa H$), we find that:
 \begin{equation}
\tau =  \zeta n_\mathrm{g} m_\mathrm{p} \kappa H = \zeta_\mathrm{atomic} ({m_\mathrm{c}\over \cancel{m_\mathrm{p}}}) n_\mathrm{g} \cancel{m_\mathrm{p} }\kappa H = \zeta_\mathrm{atomic} m_\mathrm{c} n_\mathrm{g} \kappa H.
\end{equation}
 This makes the porous optical depth:
\begin{equation}
\tau_\mathrm{p} = \zeta_\mathrm{atomic} m_\mathrm{c} n_\mathrm{g} \kappa_\mathrm{p} H_\mathrm{p}.
\end{equation}
 Next, we use the expression for the gas optical depth $\tau_\mathrm{g}= n_\mathrm{g} m_\mathrm{g} \kappa_\mathrm{g} H_\mathrm{g}$, so that we can rewrite the ratio of porous optical depth to gas optical depth as:

\begin{equation}
{\tau_\mathrm{p}\over \tau_g}= \frac{\zeta_\mathrm{atomic} m_\mathrm{c} \cancel{n_\mathrm{g}} \kappa_\mathrm{p} H_\mathrm{p}} {\cancel{n_\mathrm{g}} m_\mathrm{g} \kappa_\mathrm{g} H_\mathrm{g}}. 
\end{equation}
If we substitute in Equation \ref{H_por_rh} ($H_\mathrm{p} = H_\mathrm{g} {r_\mathrm{H}\over r_\mathrm{p}} (1-\phi)^{-1}$),

\begin{equation}
{\tau_\mathrm{p}\over \tau_g}= \frac{\zeta_\mathrm{atomic} m_\mathrm{c} \kappa_\mathrm{p} \cancel{H_\mathrm{g}} {r_\mathrm{H}\over r_\mathrm{p}} (1-\phi)^{-1} } {m_\mathrm{g} \kappa_\mathrm{g} \cancel{H_\mathrm{g}}}. 
\end{equation}
Rearranging, we find
\begin{equation}
{\tau_\mathrm{p}\over \tau_g} = \left({m_\mathrm{c}\over m_\mathrm{g}}\right) \,\, \left(\zeta_\mathrm{atomic} (1-\phi)^{-1}\right)   \left({r_\mathrm{p} \over r_\mathrm{H}}\right)^{-1}  \left({\kappa_\mathrm{p}\over \kappa_\mathrm{g}}\right)
\end{equation}
If we insert the expression for opacity of a simple opaque particle (Equation \ref{eq_opacity}) and the porous particle radius in terms of its solid counterpart (Equation \ref{same_mass_r}), we get 

\begin{equation}
{\tau_\mathrm{p}\over \tau_g} = \left({m_\mathrm{c}\over m_\mathrm{g}}\right) \,\, \left(\zeta_\mathrm{atomic} (1-\phi)^{-1}\right)   \left({r_\mathrm{s}(1-\phi)^{-{1\over3}} \over r_\mathrm{H}}\right)^{-1}  \left({\kappa_\mathrm{s} (1-\phi)^{-{2\over3}}\over \kappa_\mathrm{g}}\right)
\end{equation}
which simplifies to

\begin{equation}
{\tau_\mathrm{p}\over \tau_g}= \left({m_\mathrm{c}\over m_\mathrm{g}}\right) \,\, \left(\zeta_\mathrm{atomic} (1-\phi)^{-{4\over3}}\right)   \left({r_\mathrm{H} \over r_\mathrm{s}}\right)  \left({\kappa_\mathrm{s}\over \kappa_\mathrm{g}}\right).
\end{equation}


 
If {aerosol} particles are porous, the abundance appears to be ``enhanced'' by porosity, giving an effective abundance over the atomic abundance of the condensate in the atmosphere $\left[(1-\phi)^{-1} \zeta_\mathrm{atomic}\right]$, where $\zeta_\mathrm{atomic}$ could be set to its cosmic value ($\zeta^{*}$), where appropriate. 

\begin{figure*}[h!]
\begin{center}
\includegraphics[width=0.99\textwidth]{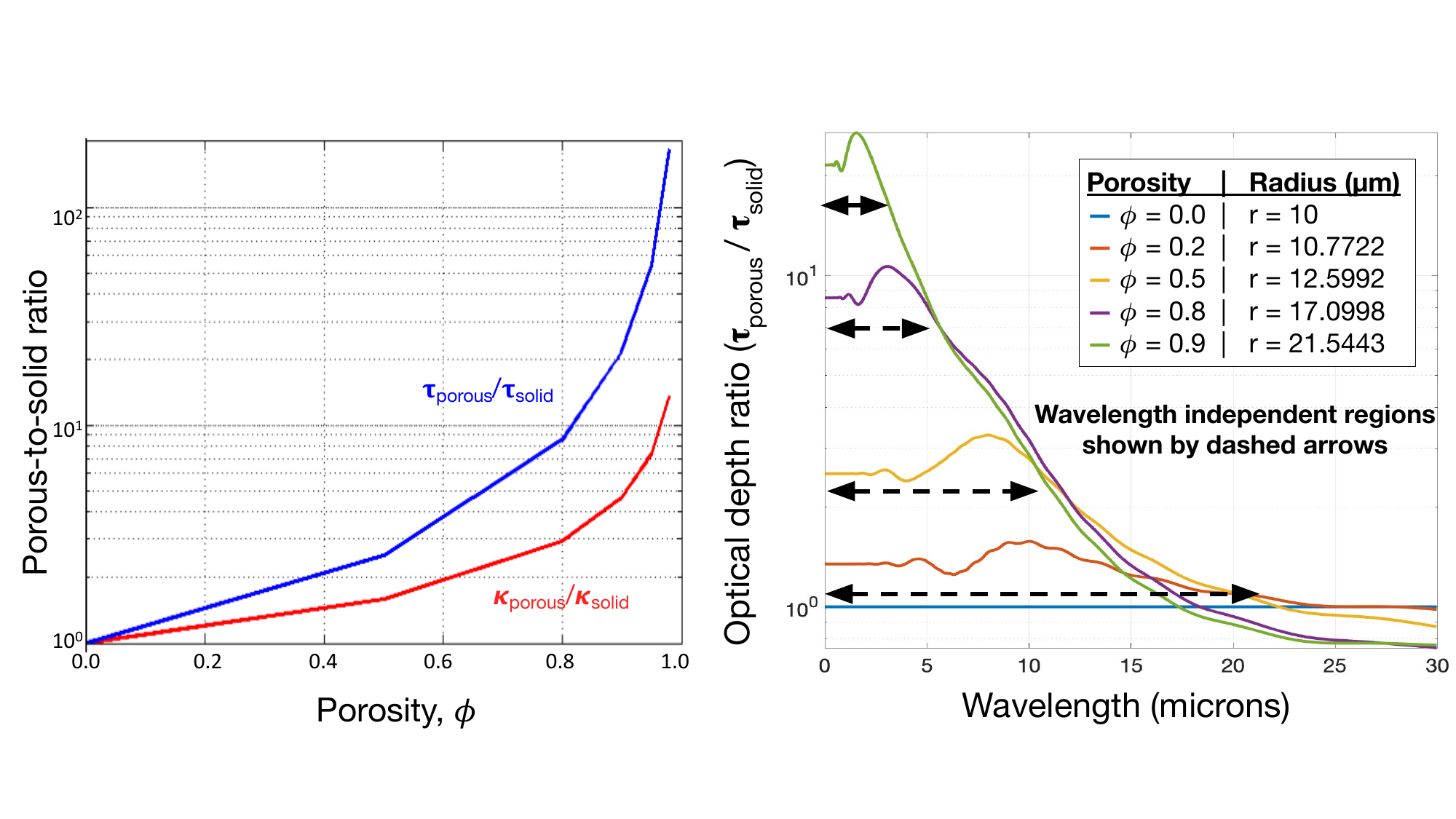}
\caption{The enhancement of optical depth and opacity due to porosity versus fall off at shorter wavelengths (i.e, the tension between two competing factors). Left: ratios of porous to solid particle vertical optical depth $\tau$ (blue) and opacity $\kappa$ (red), from Equations \ref{eq_opacity} and \ref{eq_tau}. For these calculations, the porous and solid particles have the same mass and are in the wavelength independent regime with the assumption $\sigma_\mathrm{ext}=\pi r^{2}$. Right: the ratio of vertical optical depth for porous particles to solid particles with a radius of 10 microns. The porosity is increased while keeping the mass the same as the solid 10 micron particle equivalent, resulting in larger and larger effective radii. These calculations are done using Equation \ref{general_tau_ratio} with no assumptions for $\sigma_\mathrm{ext}$. The scattering efficiencies are calculated using a Mie code.}
\label{opacity}
\end{center}
\end{figure*}

While this enhancement is promising in that it could explain flat transmission spectra by invoking porous, high altitude aerosols shrouding many planets, caution and care must be taken in how porosity is used to retrieve particle properties. The cautionary tale is demonstrated when considering the maximum wavelength of the wavelength independent regime: as porosity increases, the particles become more ``transparent" and their opacity rolls off from the wavelength independent regime at shorter wavelengths, as shown in {\bf Figure \ref{opacity}}, right panel, and also discussed in \citet{Cuzzi2014}.

\subsection{Particle property parameter space in the wavelength independent regime}  \label{section:param_space}

As shown in the last sections, the extinction efficiency and the wavelength independent regime are functions of the refractive index, which is directly related to the porosity of the particle. When porosity increases, the overall refractive index decreases, which means the optical phase shift, $\varrho$, through the center of the particle decreases (see Section \ref{section:solid}, also \citealt{hulst1981}, Chapter 11.23, Figure 33). This then results in a lower $Q_\mathrm{ext}$. Since the phase shift traces the extinction efficiency through the transition between the Rayleigh regime and the wavelength independent regime, we can use it to set limits on the grain size, mass, and porosity necessary to flatten a spectrum up to a maximum wavelength ($\lambda_\mathrm{max}$). The criterion for wavelength-independent extinction is that the phase shift of a ray passing through the center of a porous particle with size parameter $x=2\pi r_\mathrm{p} / \lambda$ needs to satisfy Equation \ref{eq:phaseshift} such that $\varrho = 2x(n_\mathrm{r_\mathrm{p}}-1) \gtrsim \varrho^{*}\sim3$, where the onset of the wavelength independent regime at $\varrho=\varrho^{*}$ can be observed from a comparison of $Q_\mathrm{ext}$ vs. $\varrho$ as shown in Figure \ref{QE}, right. The real component of the refractive index of a porous particle ($n_\mathrm{r_\mathrm{p}}$) is a function of its porosity. For simplicity, if we use volume mixing theory for the refractive index of porous particles, we can write 

\begin{equation}
n_\mathrm{r_\mathrm{p}}=1+\mbox{ff}(n_\mathrm{r_\mathrm{s}}-1)=1+(1-\phi)(n_\mathrm{r_\mathrm{s}}-1)
\end{equation} as a function of the solid refractive index ($n_\mathrm{r_\mathrm{s}}$) and filling factor($\mbox{ff}$) or porosity ($\phi=1-\mbox{ff}$). Using the porous refractive index and the size parameter, we get the following expression for the phase shift criterion as a function of size, porosity, and refractive index:

\begin{equation}
{4\pi r_\mathrm{p} \over \lambda_\mathrm{max}}(n_\mathrm{r_\mathrm{p}}-1) ={4\pi r_\mathrm{p} \over \lambda_\mathrm{max}} (1-\phi) (n_\mathrm{r_\mathrm{s}}-1) \gtrsim \varrho^{*}, \mbox{where $\varrho^{*} \sim 3$}
\label{Phase_criterion}
\end{equation}

\begin{figure}[!ht]
\hspace*{\fill}
\begin{minipage}[c]{0.5\textwidth}
\resizebox{\linewidth}{!}{\includegraphics{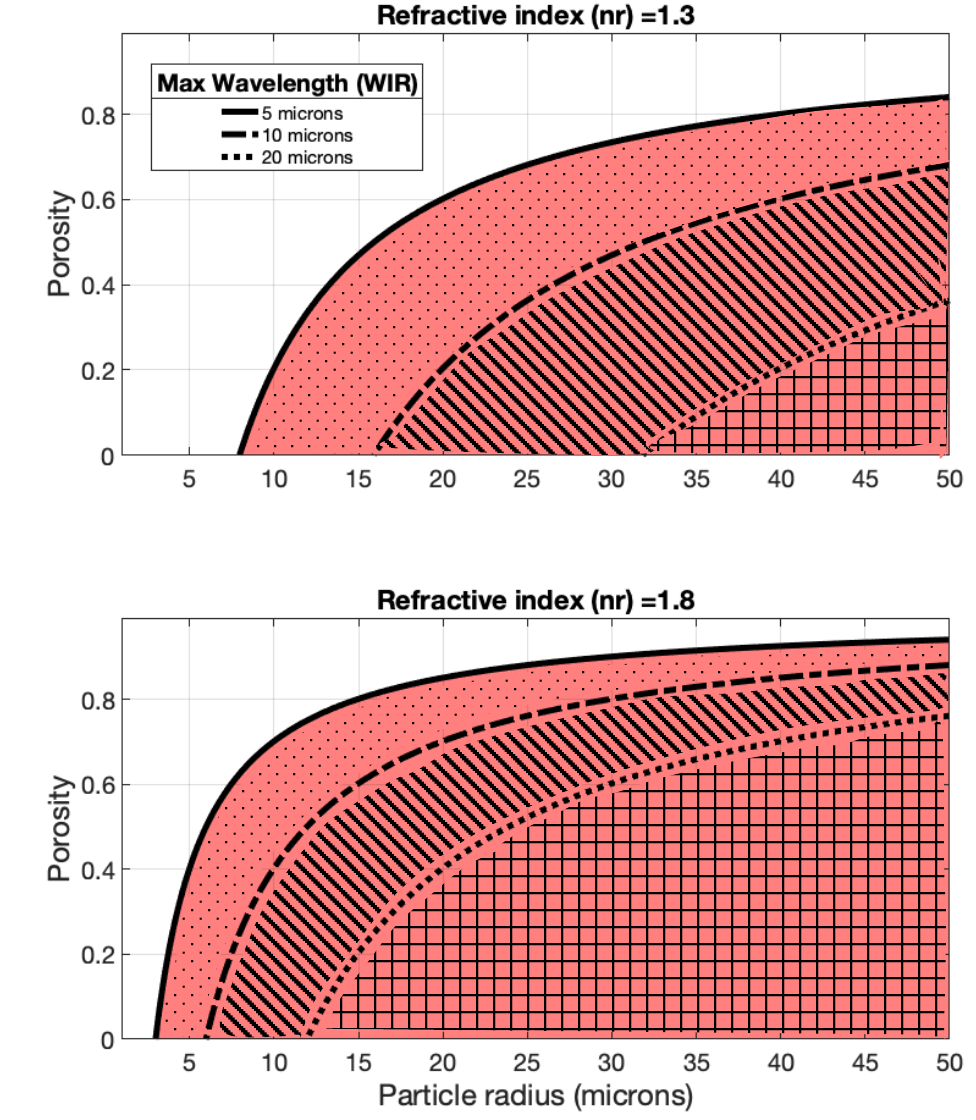}}
\end{minipage}
\begin{minipage}[c]{0.45\textwidth}
\caption{Porous particle parameter space regimes using the optical phase shift criteria from Equation \ref{Phase_criterion}. The wavelength-independent regime is shown by the shaded pink region bounded by the various black line styles and depicted by different shading styles, which represent the maximum wavelength up to which point the spectrum is wavelength independent. The maximum wavelength shown here ranges from 5 to 10 to  20 microns. The calculation is repeated for two different refractive indices shown in the top (refractive index n$_\mathrm{r}$=1.3) and bottom (n$_\mathrm{r}$=1.8)  panels, showing that increasing the refractive index increases the extent of the wavelength independent regime.}
\label{PhSp}
\end{minipage}
\end{figure}

The particle parameter space needed for wavelength independence will depend on the maximum wavelength ($\lambda_\mathrm{max}$), up to which point the spectrum remains flat ({\bf Figure \ref{PhSp}}). For example, as $\lambda_\mathrm{max}$ increases from 5 to 20  $\mu m$ (shown in {\bf Figure \ref{PhSp}} by different shaded regions), larger particles are needed, and higher porosity particles need larger refractive indices to maintain the wavelength independent regime.  

 We can also rewrite the phase space criterion in terms of the mass of a grain ($m_\mathrm{p}$), where $\rho_\mathrm{s}$ is the density of a solid particle:

\begin{equation}\label{criterion}
m_\mathrm{p} \gtrsim {{(\varrho^{*}\lambda_\mathrm{max})}^{3} \rho_\mathrm{s} \over 48 \pi ^{2}} (n_\mathrm{r_\mathrm{s}}-1)^{-3} (1-\phi)^{-2}
\end{equation}

This is the minimum particle mass necessary to flatten a spectrum up to $\lambda_\mathrm{max}$, assuming that the observation has captured $\lambda_\mathrm{max}$.  If $\lambda_\mathrm{max}$ is longer than the wavelength range captured by observations, then this is only a lower limit on $m_{\rm{p}}$ or $r_{\rm{p}}$.  This limit is only giving information on the maximum wavelength observed, and it may be that the particles are larger if $\lambda_\mathrm{max}$ is longer. As an example, consider that the phase shift criterion is met at $\varrho^{*}$ = 3. For the silicate enstatite, MgSiO$_3$, with  $\rho_\mathrm{s}$ = 4.1 g cm$^{-3}$ and n$_\mathrm{r_\mathrm{s}}$ = 1.47, and using a porosity $\phi$ = 0.9, the minimum particle mass to cause a flat spectrum up to 10 microns is $\sim$0.2 $\mu$g. On the other hand, if the particle is solid (i.e, is non-porous, $\phi$ = 0), then the minimum particle mass to flatten the spectrum to $\lambda_\mathrm{max}$ = 10 microns is two orders of magnitude smaller. 

\newpage
\section{Fractal aggregates} \label{section:fractal}

So far, we have explored the implications of homogeneous porous particles on the wavelength independent regime. A more realistic cloud particle or aerosol scenario is likely represented by aggregates composed of smaller monomers, formed from a total mass of solids ($M_\mathrm{tot}$) in a column of atmosphere with volume $V$. This allows us to treat more morphologies than just porous spheres, including potentially crystalline cloud particle structures.  In the following sections we will demonstrate the effects of fractal aggregates. We can imagine the condensates as monomers with radius and mass ($r_\mathrm{o}$, $m_\mathrm{o}$) that can stick together to form larger clusters referred to as aggregates of varying size, shape, and porosity. Depending on how many monomers are taken up in each aggregate particle, the total number density of aggregates will be 

\begin{equation}
n={(M_\mathrm{tot}/V) \over M_\mathrm{agg}} \mbox{ where $M_\mathrm{agg}$ is the individual aggregate mass defined below}.
\label{column_mass1}
\end{equation}
Fractals are defined as a collection of self-similar units called monomers with the following relationship\footnote{Here we have excluded a parameter known as the fractal pre-factor, \textit{k$_f$}, which we assume is near enough unity to ignore for our purposes. Other works \citep[e.g.,][]{tazaki2018,Ohno2020} delve more explicitly into treatments of this prefactor.} between the fractal dimension $D$ (a parameter that characterizes the piece to the whole), aggregate radius $R_\mathrm{agg}$, the number $N$ of monomers, and their size $r_\mathrm{o}$:
 
\begin{equation}
N=({R_\mathrm{agg} \over r_\mathrm{o}})^{D}.
\label{monomer_number}
\end{equation}

\begin{figure}[!ht]
\begin{center}
\includegraphics[width=0.99\textwidth]{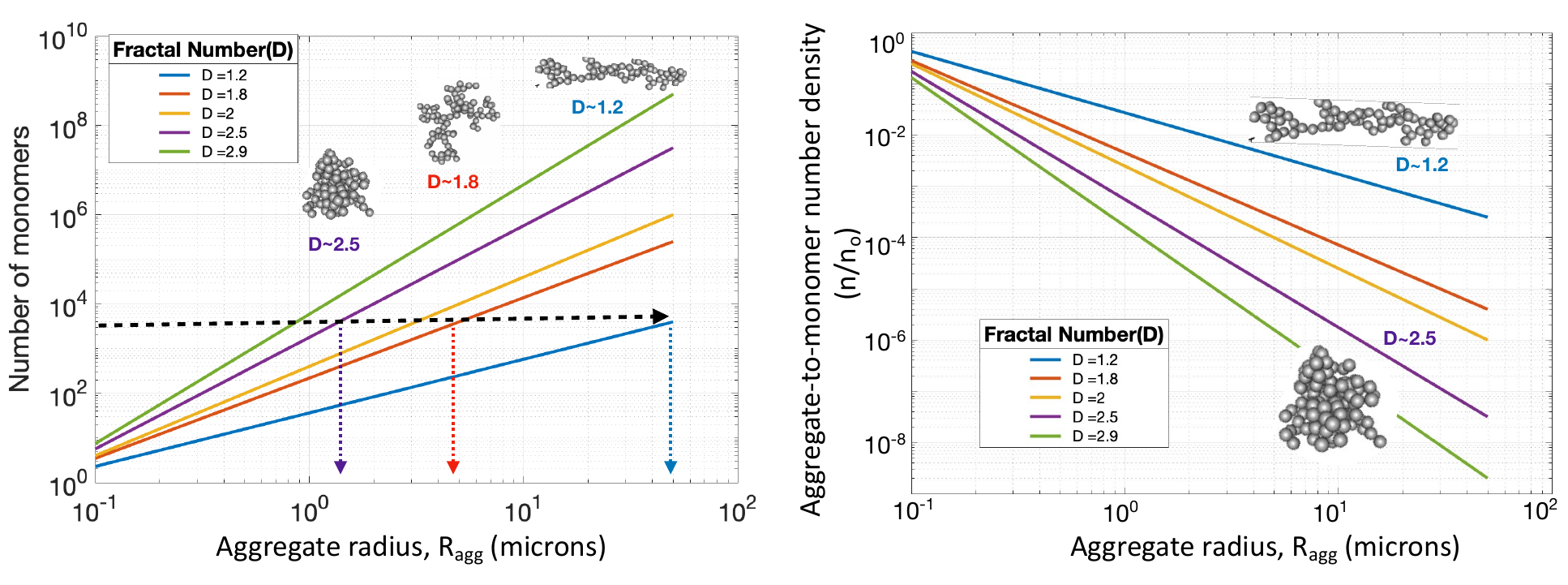}
\caption{Left: This figure shows the relationship between the aggregate radius and the number of monomers (or mass) on the y-axis. For some constant number (or mass) of monomers, indicated by the dashed black line, then the lacy aggregates ($D=1.2, 1.8$) have larger characteristic radii than compact aggregates ($D=2.9$), as indicated 
by the purple, red, and blue arrows. Right: Fractal aggregate number density to monomer number density ratio, ${n}\over{n_\mathrm{o}}$. For the same aggregate radius, there are more lacy aggregates than compact aggregates since compact aggregates have more monomers, i.e., they take up more condensate per particle.}
\label{FracAgg}
\end{center}
\vspace{0.1in}
\end{figure}

\begin{figure}[!ht]
\begin{center}
\includegraphics[width=0.99\textwidth]{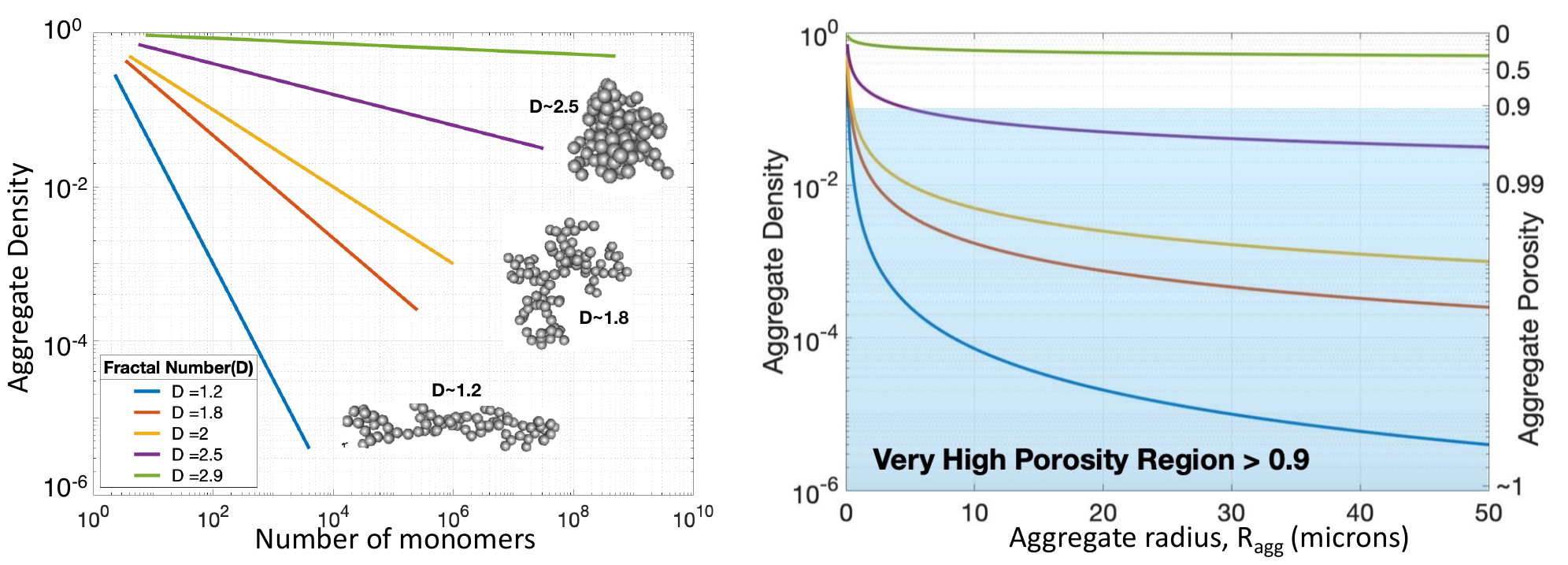}
\vspace{-0.1in}
\caption{Left: The aggregate density or filling factor as a function of the number of monomers for different aggregate types as indicated by fractal number $D$. As you add mass to an aggregate (or increase the number of monomers) the density decreases for aggregates with $D<3$. Right: Aggregate density (left y-axis) and porosity (right y-axis) as a function of the aggregate radius $R_\mathrm{agg}$. With increasing aggregate radii, the aggregate density decreases while the porosity ($1-\rho_\mathrm{agg}$) increases. The region of very high porosity is indicated by blue shading.}
\label{FracAggDensity}
\end{center}
\vspace{0.1in}
\end{figure}

Since the important parameters in question are the density and radius of the aggregate, which dictate its opacity and dynamics, it is instructive to understand how the density varies as a function of fractal dimension and radius (see {\bf Figure \ref{FracAgg}}). From Equation \ref{monomer_number} above, we can see that given an aggregate and monomer radius, the number of monomers depends on the fractal type described by the fractal dimension $D$. Assuming the same total mass of the aggregate, more compact fractals with $D>2$ have more monomers than lacy structures with $D<2$, also shown in {\bf Figure \ref{FracAgg}}. The density of a fractal ($\rho_\mathrm{agg}$) changes as you add monomers for different fractal types and is shown in {\bf Figure \ref{FracAggDensity} and Equation \ref{aggregate_mass_density}}. The important take home message from {\bf Figure \ref{FracAggDensity}} is that aggregates with $D<2.5$ and radius $R_\mathrm{agg}$ greater than $\sim$ a few 
microns have very high porosities, $\phi$ $>0.9$. The majority of spherical particle haze models \citep[e.g.,][]{MillerRicciKempton2012,morley2013,Gao2023} have particle sizes less than 10 microns and assume compact spherical shapes. This assumption leads to the higher density regime, which will affect how long particles can stay aloft, and thus the opacity of the atmosphere at these high altitudes.

The mass and density of an aggregate are described as:
\begin{equation}
M_\mathrm{agg}=N m_\mathrm{0} = m_\mathrm{0} \left({R_\mathrm{agg} \over r_\mathrm{0}}\right)^{D}, \mbox{ and  } \rho_\mathrm{agg}=\rho_\mathrm{0}\left({R_\mathrm{agg} \over r_\mathrm{0}}\right)^{D-3}
\label{aggregate_mass_density}
\end{equation}

With porosity ($\phi$) defined as $\mbox{ff}=1-\phi$, aggregate porosity and the filling factor (ff) can be derived as follows:

\begin{equation}
\phi=1-\mbox{ff}=1-N\left({r_\mathrm{o} \over R_\mathrm{agg}}\right)^{3}= 1- \left({R_\mathrm{agg} \over r_\mathrm{o}}\right)^{D-3}, \mbox{and  } \rho_\mathrm{agg}=\mbox{ff}\rho_\mathrm{0}
\label{aggregate_porosity}
\end{equation}
Note that a solid, compact particle -- with $\mbox{ff}=1$ and $\phi = 0$ -- therefore has a  fractal dimension $D = 3$.

Using the fractal properties from above, the aerosol number density $n$ can be written as:

\begin{equation}
n={ (M_\mathrm{tot}/V) \over M_\mathrm{agg}}={ (M_\mathrm{tot}/V) \over {Nm_\mathrm{o}}}=n_\mathrm{o}({r_\mathrm{o}\over {R_\mathrm{agg}}})^{D}\\
\label{column_mass2}
\end{equation}

where $n_\mathrm{o}$ is the monomer number density if the condensates in the column were present as separate monomers. In the case of fractal aggregation, the number density is always lower than $n_\mathrm{o}$.\\

\newpage
\subsection{Realistic aggregate optical properties and effects in the wavelength independent regime}
The purpose of this section is to demonstrate the behavior of realistic aggregates. 
We use the Discrete Dipole Approximation (DDA) to model aggregates and calculate their scattering properties as a function of wavelength and refractive indices (for recent other explanations of DDA, see \citealt{lodge2024}). Our understanding so far has been based on Mie-EMT models, where we showed that the wavelength independent regime has a shorter wavelength range for porous particles than for solid particles of the same composition. 

\begin{figure*}[h!]
\begin{center}
\includegraphics[trim={6.7cm 0cm 5cm 0cm},clip,width=0.57\textwidth]{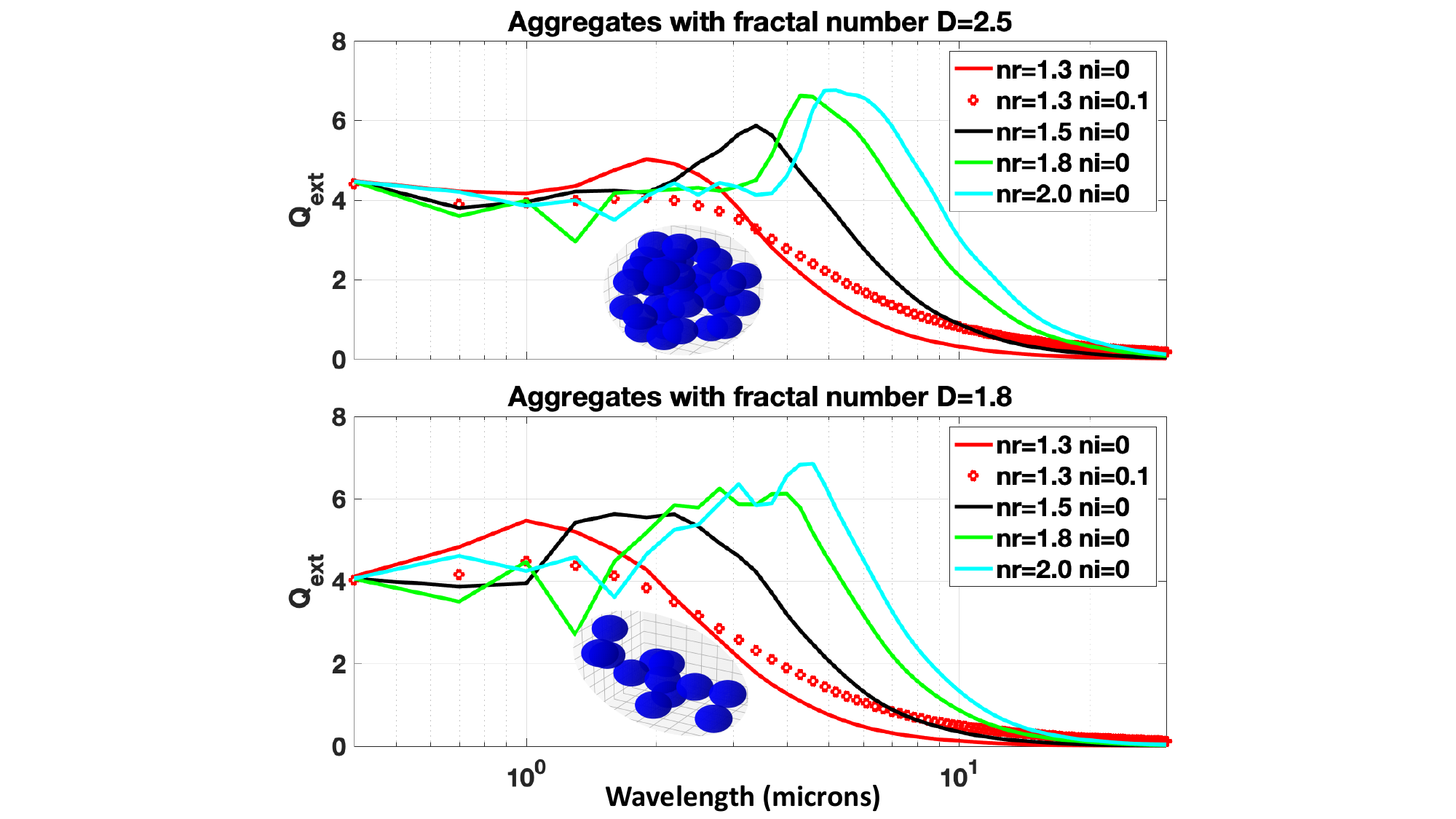}
\label{imag_index}
\end{center}
\caption{Extinction efficiency Q$_{\rm{ext}}$ vs. wavelength for DDA aggregate runs of two fractal types (top: compact; bottom: lacy) with varied refractive indices. All runs were for an imaginary refractive index of $n_{\rm{i}}$ = 0, except for the red diamonds, corresponding to an imaginary refractive index of $n_{\rm{i}}$ = 0.1. The imaginary component of the refractive index changes the slope of the Q$_{\rm{ext}}$ vs. wavelength line, as seen comparing the solid red line with the same real refractive index. Larger fractal number $D$ has a more extended wavelength-independent regime.}
\label{DDAimg}
\end{figure*}

\begin{figure*}[h!]
\begin{center}
\includegraphics[width=0.99\textwidth]{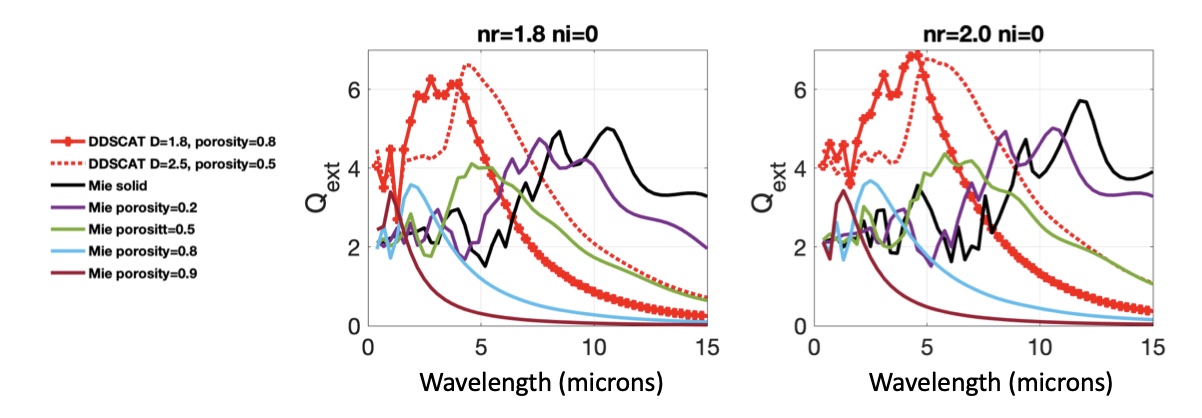}
\label{PhaseSpace}
\end{center}
\caption{Extinction efficiency vs wavelength for aggregates computed using either the Mie-EMT method or the more realistic DDSCAT technique . Each panel shows the comparison for purely scattering particles with a different real refractive index (all $n_\mathrm{i}=0$). DDSCAT aggregates (red curves) have a higher Q$_\mathrm{ext}$ at shorter wavelengths than homogeneous Mie-EMT spheres. When the real refractive index is increased, the peak of Q$_\mathrm{ext}$ increases and shifts to longer wavelengths for all particles.}
\label{DDA}
\end{figure*}

In order to extend the wavelength independent regime for very porous particles, the simple Mie-EMT models of Section \ref{section:param_space} suggest that refractive indices must increase (hence the inference of very fluffy, high refractive index particles up high in the atmosphere to explain flat exoplanet transmission spectra). These DDA calculations are meant to test this result that was based on the simplistic Mie model. {\bf Figure \ref{DDAimg} and \ref{DDA}} show a set of runs for two aggregates with porosities 0.5 (compact, D = 2.5) and 0.8 (open, D = 1.8). Scattering calculations were performed with the DDSCAT code \citep{Draine1994} as a function of wavelength and refractive index. 

Figure \ref{DDAimg} demonstrates how more lacy particles with increasing refractive index have narrower wavelength independent regimes compared to compact particles, with a bluer peak $Q_{\rm{ext}}$. The figure also shows that the imaginary component of the refractive index changes the slope of the extinction curve at optical and near-infrared wavelengths, as could also flatten exoplanet transmission spectra as seen in Figure \ref{Hazetransit}. The nonzero $n_{\rm{i}}$ also damps the amplitude of the first resonant peak in $Q_{\rm{ext}}$. 

Figure \ref{DDA} shows that the wavelength independent range for the compact DDA aggregate (porosity $\phi$ = 0.5, dashed red curve) is closer to the Mie-EMT homogeneous porous particle with the same porosity (shown in green in  Figure \ref{DDA}) than the open aggregate with higher porosity. At higher porosities ($0.8$, $0.9$), the realistic DDA aggregate has a more extended wavelength-independent regime than the homogeneous case. 

Also, this highlights the inherent degeneracy between number density, mass, porosity, and compactness of a given particle size and shape to explain observations. For example, an atmospheric retrieval using the Mie-EMT model would require a more compact particle in order to maintain a certain wavelength-independent regime.  In this case, the particle mass might be overestimated, as compact particles require more particles overall to achieve the needed $Q_\mathrm{ext}$, as has been also observed when using multiple approaches to model the icy ring grains of Saturn \citep{Vahidinia2011}.

\newpage
\newpage
\section{Upper atmosphere aerosol transport: Drag force and particle-gas regimes}
\begin{figure*}[h!]
\begin{center}
\includegraphics[angle=0,width=.75\textwidth]{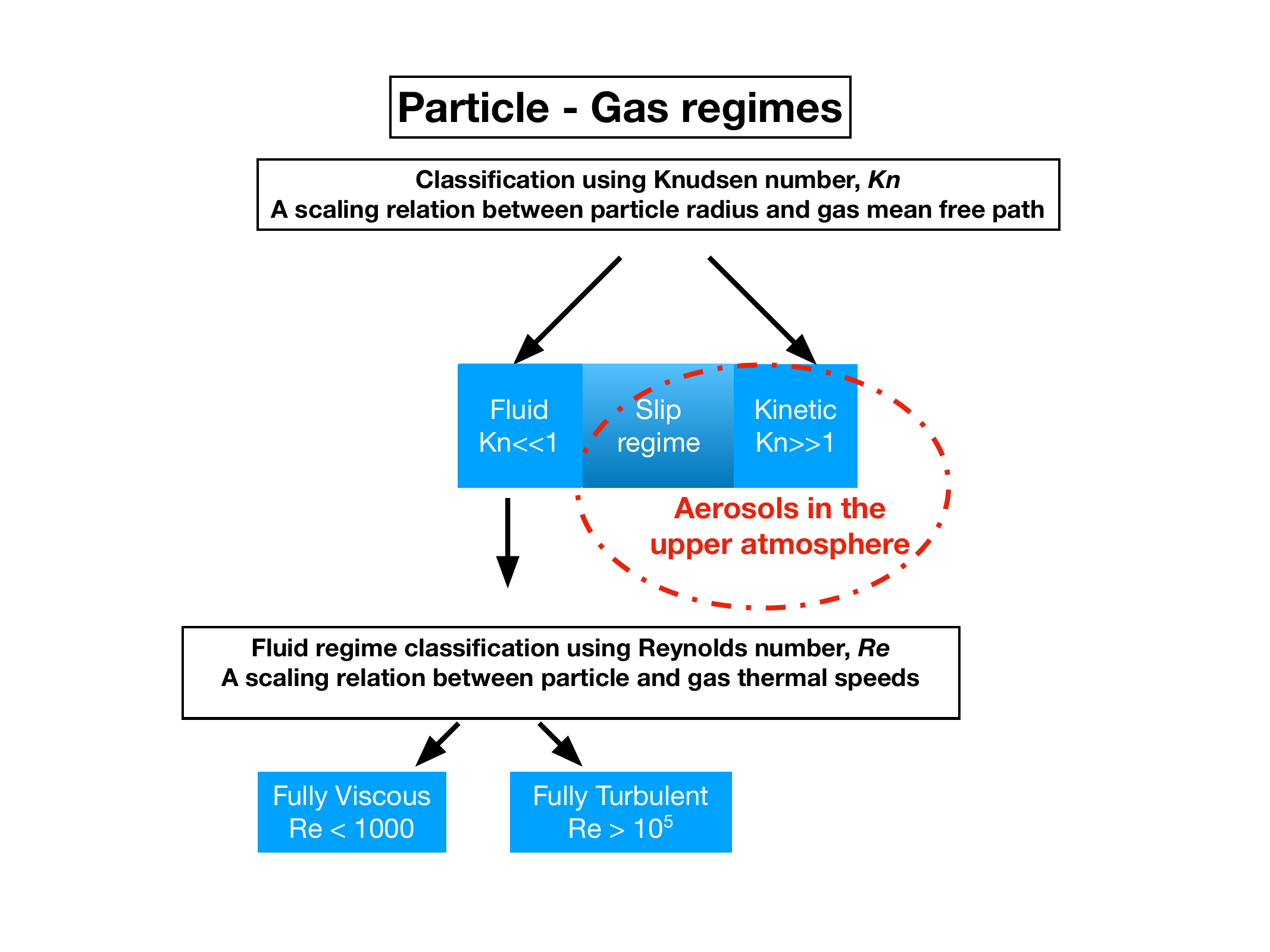}
\vspace{-0.1in}
\caption{Diagram showing the particle-gas regimes starting with the Knudsen number, $K\!n$, which is the ratio of the particle radius to the gas mean free path. This Knudsen number distinguishes between the \textbf{fluid} regime where the particles are small compared with the gas mean free path, and the \textbf{kinetic} regime where the particles are large compared with the gas mean free path. The fluid regime can be further broken down by the Reynolds number, $Re$, which distinguishes between a \textbf{turbulent} (high $Re$) flow past the particle and in its wake dominated by eddies and vortices and a \textbf{laminar} (low $Re$) flow dominated by viscous forces.}
\label{atmos_regimes}
\end{center}
\end{figure*}

Aerosol transport in the upper atmosphere depends on the drag force imparted on the particles, which affects the particle residence time throughout the atmosphere. 
Since the drag force depends on the atmospheric density and relative velocity of the gas and particle, there are different physics at play depending on the atmospheric structure. The interaction of the particle and the gas is determined by the relative size of the particle and mean free path of the gas (see {\bf Figure \ref{atmos_regimes}}).

\begin{figure*}[h!]
\begin{center}
\includegraphics[width=0.8\textwidth]{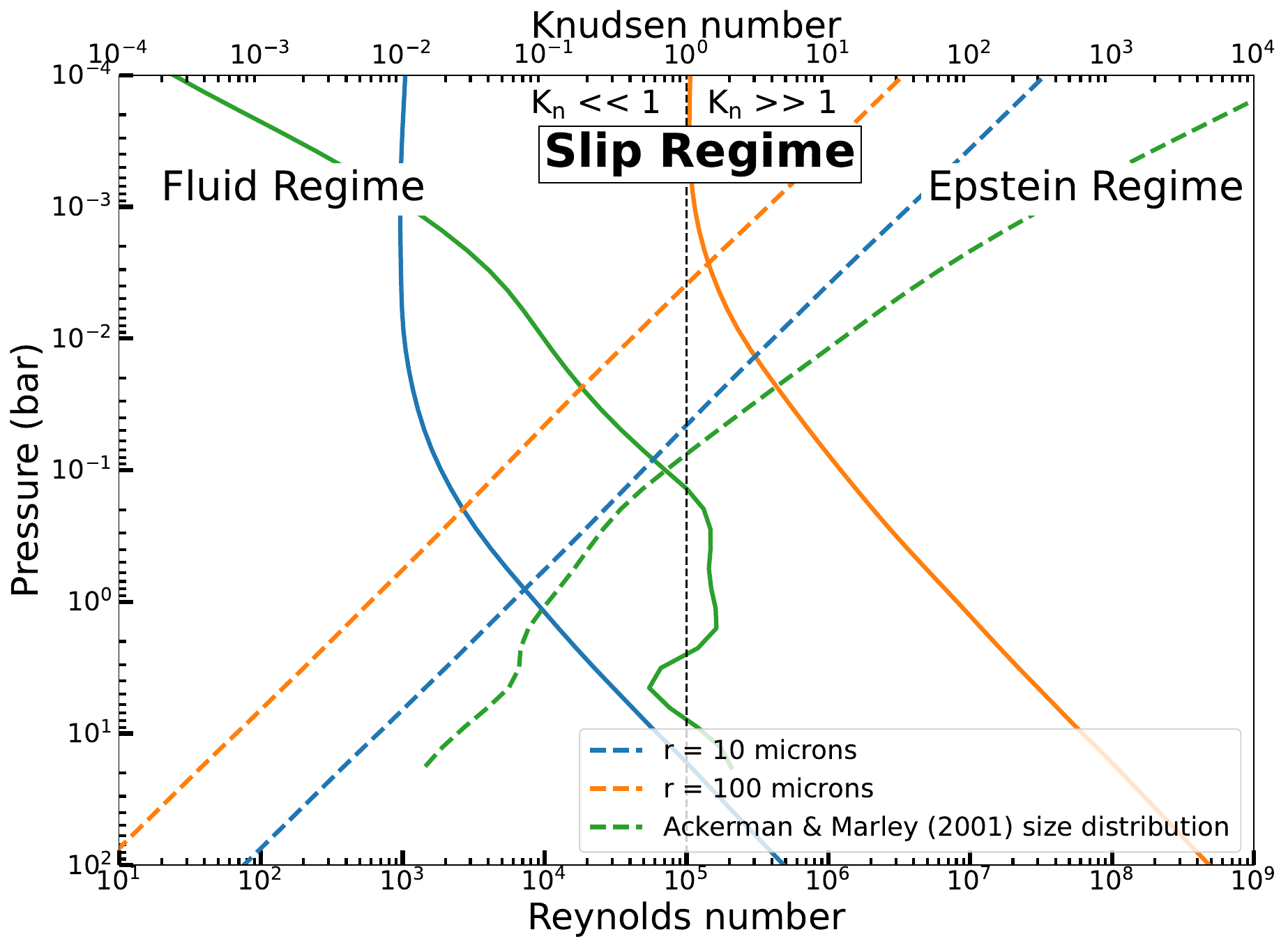}
\caption{Knudsen and Reynolds numbers for various particle sizes as a function of pressure. The particle-gas regime delineation can start with the Knudsen number, which bifurcates the space into the fluid (also referred to as Stokes) and kinetic (also referred to as Epstein or free molecular) regimes. Both Knudsen (dashed) and Reynolds (solid) numbers are shown in log space. From these calculations, it is evident that the upper atmosphere -- which we can take to be pressures above $10^{-1}$ bar -- is in the kinetic regime for particle sizes of interest.}
\label{atmos_Kn_Re}
\end{center}
\end{figure*}

If the particle size is large compared to the gas mean free path (i.e., the fluid regime), then the gas imparts fluid pressure on the particle and the frictional force can be calculated using fluid pressure from Bernoulli's laws. If the particle size is small compared to the gas mean free path (i.e., the kinetic regime, sometimes also called the Epstein regime or the free molecular regime), then the drag force is the collective effect of collisions of individual molecules in the gas. The bifurcation of the regimes can generally be characterized by the Knudsen number, $K\!n$, which differentiates between the fluid regime and the kinetic or free molecular regime. The Knudsen number is given by:

\begin{equation}
    K\!n={\lambda \over r}
\end{equation}
where $\lambda$ is the molecular mean free path and $r$ is radius of the particle. These two regimes are shown in {\bf Figure \ref{atmos_Kn_Re}}. The region between $0.4\ll K\!n \ll20$ is the transition region between the fluid regime, where the particle ``sees" the gas as a continuous fluid, and the kinetic regime, where the particle ``sees'' individual collisions by individual gas molecules. 
These regimes -- and where they begin and end in the atmosphere -- provide context for understanding the transport scenarios of aerosols, especially for larger aggregate particles at high altitudes. 

There are various pathways that particles can take to reach the kinds of high altitudes relevant to the transition between these flow regimes. For example, one scenario is that large aerosols can be formed at high altitude via UV bombardment of simple organic molecules, which creates tiny monomers of refractory organics, which can subsequently coagulate to form aggregates and settle (e.g., photochemical haze particles; \citealt{Adams2019}). Another scenario is that more refractory condensate seeds can be transported via turbulent mixing and advective gas flows from deep in the atmosphere and then make their way to higher altitudes via Brownian diffusion while serving as sticking sites for other condensates to form aggregates (discussed further in Section \ref{sec:brownian}). In either case, the \textit{stopping time} plays a key role in the aerosols' residence time in the upper atmosphere.  

\subsection{Aggregate stopping time and terminal velocity} \label{section:stop_time}

The time a particle spends at a particular altitude depends on its terminal velocity, which is the velocity where the gravitational force is balanced by the drag force -- i.e., there is no net acceleration or change in velocity. Therefore, the more frictional force, the faster the particle stops accelerating through the atmosphere. The time it takes to alter a particle's momentum until it reaches its terminal velocity is the stopping time ($t_\mathrm{s}$), after which the particle settles down with its terminal speed. For the sake of clarity, we refer to \textit{settling time} as the time it takes to fall one atmospheric scale height at terminal velocity. Settling time is inversely proportional to the stopping time, $t_{\rm{s}}$. The faster a particle reaches terminal speed (i.e., the shorter its stopping time), the longer it takes to settle, which means the longer it will spend at high altitudes, and therefore perhaps contribute to high altitude opacity. Terminal velocity ($v_\mathrm{f}$) can be derived by setting the drag force ($F_\mathrm{drag}$) equal to the gravitational force:

\begin {equation}
F_\mathrm{drag}={\pi c_\mathrm{d} \rho_\mathrm{g} v^{2}_\mathrm{f} d^{2} \over 8 \beta } = {1\over 2}{c_\mathrm{d}\over \beta} (\pi r^{2} v^{2}_\mathrm{f} ) \rho_\mathrm{g}=mg
\end{equation}

\noindent where $c_d$ is the drag coefficient, $\rho_g$ is the gas density, $v_f$ is the fall velocity, $d$ is the particle diameter, $\beta$ is the Cunningham slip factor, $r$ is the particle radius, $m$ is the particle mass, and $g$ is the gravitational acceleration. The Reynolds number, $Re$, describes the flow as either turbulent, where $Re$ is high, or laminar, where $Re$ is low. We can express $Re$ = $6 v_f \over v_\mathrm{th} K\!n $, where  $v_\mathrm{th}$ is the thermal velocity of the gas. In the Stokes regime with no turbulence -- where we have a very low Reynolds number with $Re<1$  -- and by setting the drag coefficient $c_\mathrm{d} ={24\over R{e}}$, we can use Stokes' Law:

\begin{equation}
    F_\mathrm{drag} = 6\pi \eta r v_f
\end{equation}

\noindent where the diameter of the condensate is $d=2r$, and $\eta$ is the dynamic viscosity of the atmosphere, given by:
\begin{equation}\label{eq:eta}
    \eta={1\over3}\rho v_\mathrm{th} \lambda
\end{equation}
where $\rho$ is the density of the gas and $\lambda$ is the mean free path of the gas. Often this is presented as the kinematic or molecular viscosity, $\nu_{\rm{m}}$, which is related to the dynamic viscosity by:

\begin{equation}\label{eq:eta_kinematic}
    \nu_{\rm{m}}={\eta\over\rho}.
\end{equation}

Then, we set the Stokes drag force equal to our gravitational force to get the following expression for the fall velocity aka terminal velocity:

\begin{equation}
v_\mathrm{f}=g t_\mathrm{s}= {2 \beta g r^{2} \Delta \rho \over 9 \eta},\,\ 
\label{fall}
\end{equation}

\noindent where we have substituted in the particle density and volume for its mass, $\Delta \rho$ is the difference between the densities of condensate and atmosphere, and the Cunningham slip factor is defined as $\beta=(1+1.26K\!n)$ (a more general expression can be found in e.g., \citealt{Ohno2020}). The Cunningham factor bridges to the Epstein (aka kinetic or molecular) regime where the gas molecules are bouncing off the particle as point sources. In the Epstein regime, where $K\!n={\lambda \over r} \gg 1$ (again where $\lambda$ is the mean free path), the Cunningham slip factor becomes:

\begin{equation}
\beta = 1+1.26K\!n=1+1.26({\lambda \over r}) \sim {\lambda \over r}, \,\,\, \mbox{for} \,\,K\!n={\lambda \over r}\gg1.
\label{slipfactor}
\end{equation}

Then, we have a fall velocity expression for the Epstein or kinetic regime:

\begin{equation}
v_\mathrm{f}=g t_\mathrm{s} = {2 ({\lambda \over r}) g r^{2} \Delta \rho \over 9 \eta} = {2 \lambda g \Delta \rho r \over 9 ({1\over3} \rho v_\mathrm{th} \lambda)} = {2\over 3} {g \Delta \rho r \over \rho v_\mathrm{th}}\,\
\end{equation}

Stopping time ($t_\mathrm{s}$) in the kinetic regime can be expressed in terms of aggregate radius ($R_\mathrm{agg}$), and density ($\rho_\mathrm{agg}$), and plugging in for those variables from Equations \ref{monomer_number} - \ref{aggregate_porosity}:

\begin{equation}
t_\mathrm{s_\mathrm{agg}}={ R_\mathrm{agg} \rho_\mathrm{agg} \over c \rho_\mathrm{g}} =  t_\mathrm{o}\left({R_\mathrm{agg}\over r_\mathrm{o}}\right)^{D-2}=t_\mathrm{o}\left(1-\phi \right)^{D-2 \over D-3}
\label{stop_aggs31}
\end{equation}
where the monomer stopping time is defined as $t_\mathrm{o}= {\rho_\mathrm{o} r_\mathrm{o} \over c \rho_\mathrm{g}}$, if we set the speed of sound \textit{c} equal to the thermal velocity of the gas \textit{v$_\mathrm{th}$}, and the density of the solid monomer $\rho_o$ $\gg$ the density of the gas $\rho_g$. In order to see how the fractal type changes the stopping time, it is instructive to divide the aggregate stopping time ($t_\mathrm{s_\mathrm{agg}}$) by the stopping time for a solid spherical aggregate with the same number of monomers or mass. Thus, we're comparing an aggregate with radius $R_\mathrm{agg}=r_\mathrm{o}N^{1\over D}$ to a solid spherical cluster (i.e., \textit{D} = 3) with the same number of monomers with an outer radius defined as $R_\mathrm{solid}=r_\mathrm{o}N^{1\over3}$. The stopping time ratio for these particles is defined as:

\begin{equation}
{t_\mathrm{s_\mathrm{agg}}\over t_\mathrm{s_\mathrm{compact}}}=\left({R_\mathrm{agg}\over r_\mathrm{o}}\right)^{2D-6 \over 3}
\label{stop_aggs32}
\end{equation}

\begin{figure}[!ht]
\hspace*{\fill}
\begin{center}
\includegraphics[width=0.6\textwidth]{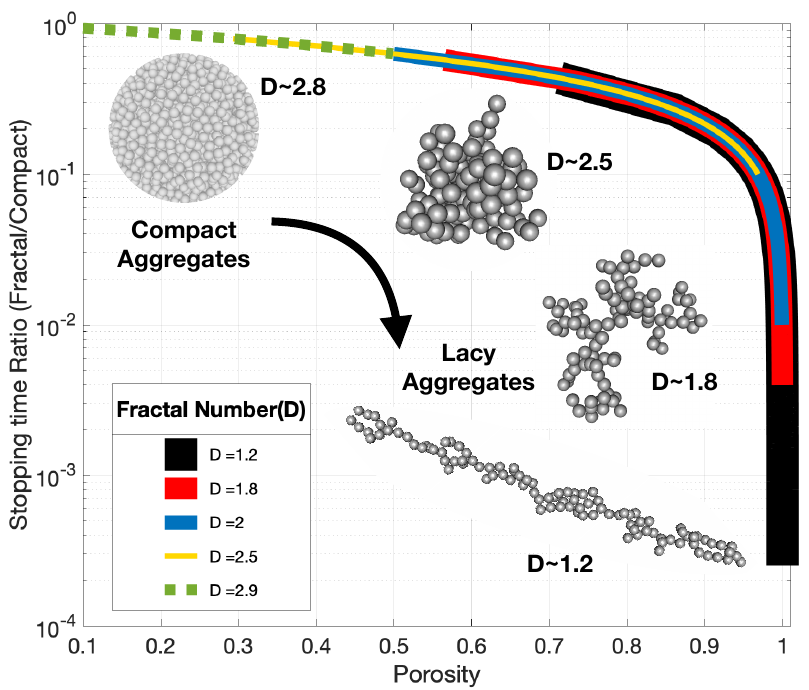}
\end{center}
\caption{Stopping time ratio for fractal aggregates compared to solid spherical aggregates. The compact case can crudely be assumed to be similar to the homogeneous porous particles considered in prior sections. In this figure, the stopping time is the time it takes for the particle to come to equilibrium with the gas. Thus, the shorter the stopping time, the quicker a particle reaches its (smaller) terminal velocity (Equation \ref{fall}), and the longer its residence time in the upper atmosphere.}
\label{stop}
\end{figure}

This ratio is shown in {\bf Figure \ref{stop}} for different fractal aggregates where the more compact particles have longer stopping times compared to lacy aggregates. Thus, the more compact particles fall further in the atmosphere to lower altitudes. There are orders of magnitude differences in settling time between the compact aggregate and lacy aggregates -- meaning the lacy aggregates spend the most time at higher altitudes.


\newpage
\subsection{Stopping time, optical phase shift, and mass per unit area}

Fractal aggregates can have important implications for microphysical modeling and atmospheric conditions that would allow certain fractal growth or types. Different fractal types -- that grow through different means -- can have the same porosity at different sizes, so considering porosity alone may not be fully illuminating. For example, fractals can grow via cluster–cluster aggregation (CCA), ballistic-CCA (BCCA), ballistic particle-cluster aggregration (BPCA), or linear chain growth \citep[see, e.g., ][]{tazaki2018,Ohno2020,tazaki2021}. As aggregates grow by accruing monomers (BPCA) or accreting other aggregates (BCCA), their stopping time and optical properties depend on the different fractal dimensions produced by these different growth paths.

The aggregate stopping time ({$t_{\mathrm{s_\mathrm{agg}}}$}) and optical phase shift ($\varrho_{\mathrm{agg}}$, which is a tracer for extinction efficiency) both depend on the mass per unit cross sectional area of the aggregate ($R_{\mathrm{agg}} \rho_{\mathrm{agg}}$). To see how this comes about, first recall from Equation \ref{Phase_criterion} that for a porous particle, the optical phase shift is $\varrho ={4\pi r_\mathrm{p} \over \lambda}(1-\phi) (n_\mathrm{r_\mathrm{s}}-1)$. Replacing porosity ($\phi$) with aggregate and monomer densities ($\rho_{\rm{agg}}$ and $\rho_{\rm{o}}$, Equation \ref{aggregate_porosity}) and the solid particle refractive index ($n_\mathrm{r_\mathrm{s}}$) with the monomer refractive index ($n_\mathrm{r_\mathrm{o}}$), we have

\begin{equation}
\varrho_\mathrm{agg} = {4\pi R_\mathrm{agg}(n_\mathrm{r_\mathrm{o}}-1) \over \lambda} {\rho_\mathrm{agg} \over \rho_\mathrm{o}} 
\label{}
\end{equation}.

 \noindent Multiplying this expression by $r_\mathrm{o}/r_\mathrm{o}$ and rearranging, we see that optical phase shift is expressed as a function of the mass per unit cross sectional area ($R_\mathrm{agg} \rho_\mathrm{agg}$):

\begin{equation}
\varrho_\mathrm{agg} = {4\pi r_\mathrm{o}(n_\mathrm{r_\mathrm{o}}-1) \over \lambda} {R_\mathrm{agg}\rho_\mathrm{agg} \over \rho_\mathrm{o} r_\mathrm{o}} 
\label{Opt_Phase_Rrho}
\end{equation}.

\noindent which is just

\begin{equation}
\varrho_\mathrm{agg} = {\varrho_\mathrm{o}} {R_\mathrm{agg}\rho_\mathrm{agg} \over \rho_\mathrm{o} r_\mathrm{o}} = \varrho_\mathrm{o} \left({R_\mathrm{agg} \over r_\mathrm{o}} \right)^{D-2}
\label{Opt_Phase_Rrho2}
\end{equation}

\noindent if we substitute in Equation \ref{aggregate_mass_density} for the aggregate density $\rho_{\rm{agg}}$.

 Notice that this equation states that the optical phase shift of an aggregate will be less than the phase shift of a solid particle for D $<$ 2, that is, for lacy aggregates. To intuitively appreciate this, it is helpful to consider a light beam interacting with both a monomer and a lacy aggregate. Averaging over a given mass per unit cross sectional area, the light beam will always interact with the monomer. However, the light beam encountering the lacy aggregate must be averaged over many random orientations of the long, lacy structure. Many of these orientations will then involve the light beam passing through the medium rather than any monomer that makes up the aggregate. Averaging over these random orientations will thus result in a lower phase shift than for a solid monomer.

From Equation \ref{Opt_Phase_Rrho}, we see that the mass per unit cross sectional area $R_\mathrm{agg} \rho_\mathrm{agg}$ emerges as:

\begin{equation}
R_\mathrm{agg} \rho_\mathrm{agg}= \rho_\mathrm{o} r_\mathrm{o} \left({R_\mathrm{agg} \over r_\mathrm{o}} \right)^{D-2}
\label{R_Rho}.
\end{equation}

\begin{figure}[!ht]
\centering
\includegraphics[width=0.4\textwidth]{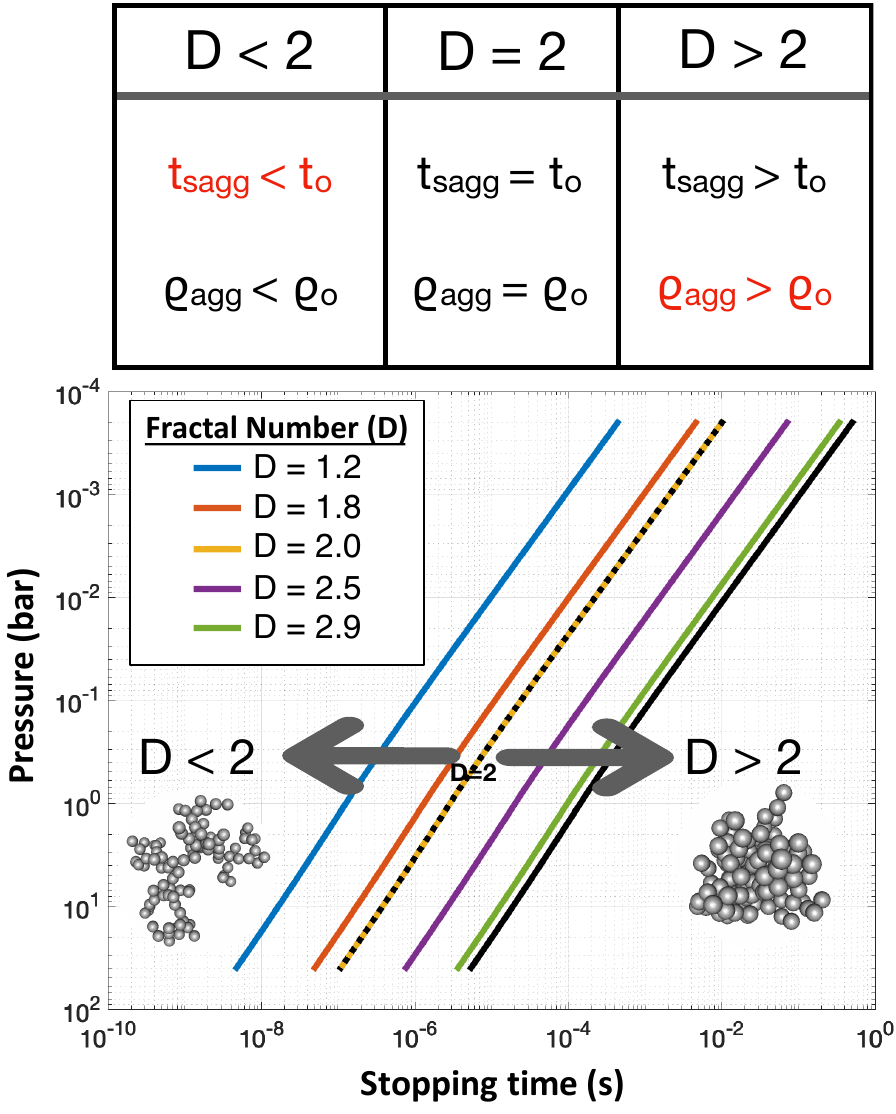}
\caption{(Top): Fractal domain chart, where the aggregate stopping time and optical phase shifts relative to the monomer's values bifurcate at $D=2$. (Bottom): Epstein regime stopping time for aggregates (colors) and solid particles (black) in the atmosphere of a planet with temperature $T_\mathrm{eq}=1600K$ and gravity $g=1000 m/s^{2}$. The aggregates with various fractal numbers are shown in different colors, and all have an outer radius of $R_\mathrm{agg}=5 \mu m$ and monomer radius $r_\mathrm{o}=100 nm$. A solid particle with the same outer radius of the aggregates is shown in solid black, and a solid particle with the monomer radius is shown as a dashed black line. The stopping time for the aggregate with fractal number $D=2$ falls along the monomer stopping time, showing the bifurcation (yellow and dashed black lines), with the lacy $D<2$ aggregate stopping times less than the monomer's, and compact $D>2$ aggregate stopping times greater than the monomer's.}
\label{stop_Ddomain}
\end{figure}

\noindent Depending on aggregate type, the mass per unit area has important implications for the stopping time and optical phase shift.  {\bf Figure \ref{stop_Ddomain}} compares three types of aggregates based on {\bf Equation \ref{R_Rho}}: lacy with $D<2$, more filled in with $D=2$, and compact with $D>2$. Lacy aggregates with fractal number $D<2$ have stopping times shorter than their individual monomers, which is favorable for having a longer residence time in the upper atmosphere; however, the phase shift also drops off as the aggregate outer radius grows, which means the monomers themselves have to have a large enough phase shift (via combination of size and refractive index) to be in the wavelength independent regime. A fractal number $D\sim2$ is interesting since both the stopping time and phase shift of an aggregate are equal to the monomer value, even as you increase the aggregate size. One could therefore build \textit{very large aggregates from monomers that are in the wavelength independent regime and have the same stopping time as the monomer}. Aggregates with fractal dimension $D>2$ are compact, and their stopping time and phase shift increase as the aggregate grows. These particles settle faster and have much shorter residence times in the upper atmosphere.  

For flat spectra caused by wavelength independent extinction, the aggregate stopping time is in a way ``tethered" to the phase shift, since some minimum phase shift ($\varrho^{*}$) defines the onset of the wavelength independent regime. Ideally, one would like to have the stopping time of a $D<2$ fractal, with the phase shift of a $D>2$ fractal, for aerosols in the upper atmosphere to cause flat spectra (as highlighted in red in the upper panel chart in {\bf Figure \ref{stop_Ddomain}}).  However, we can see that those domains do not overlap. Thus, if aggregates are the culprit for flat exoplanet transmission spectra, then the ones with fractal number $D\leq2$ (with long upper atmospheric residence times) made up of monomers that are large enough and with high enough refractive indices to be in the wavelength independent regime for a given $\lambda_\mathrm{max}$ are most plausible.  \citet{Cuzzi2014}, \citet{Adams2019}, and others discuss the compositional effects that may result in high refractive index particles. The composition of these particles is therefore a critical component for flat spectra, \textit{in addition} to the size and morphology as we have discussed in previous sections.

\newpage
\subsection{Aerosol transport via Brownian motion}\label{sec:brownian}

We have shown that low density fractal aggregates can stay aloft longer in the upper atmosphere compared to compact particles of the same mass, but aerosol transport -- i.e., how they get to the upper atmosphere -- is still a large unknown. Different scenarios have been proposed, such as {photochemical hazes} forming in the upper atmosphere via UV bombardment and settling \citep[e.g.,][]{morley2013,Adams2019}, or condensate seeds moving upward via turbulence and serving as coagulation sites for aggregates to form \citep[e.g.,][]{Ohno2020,samra2020,Samra2022}. 

\begin{figure}[!ht]
\centering
{\includegraphics[width=0.6\textwidth]{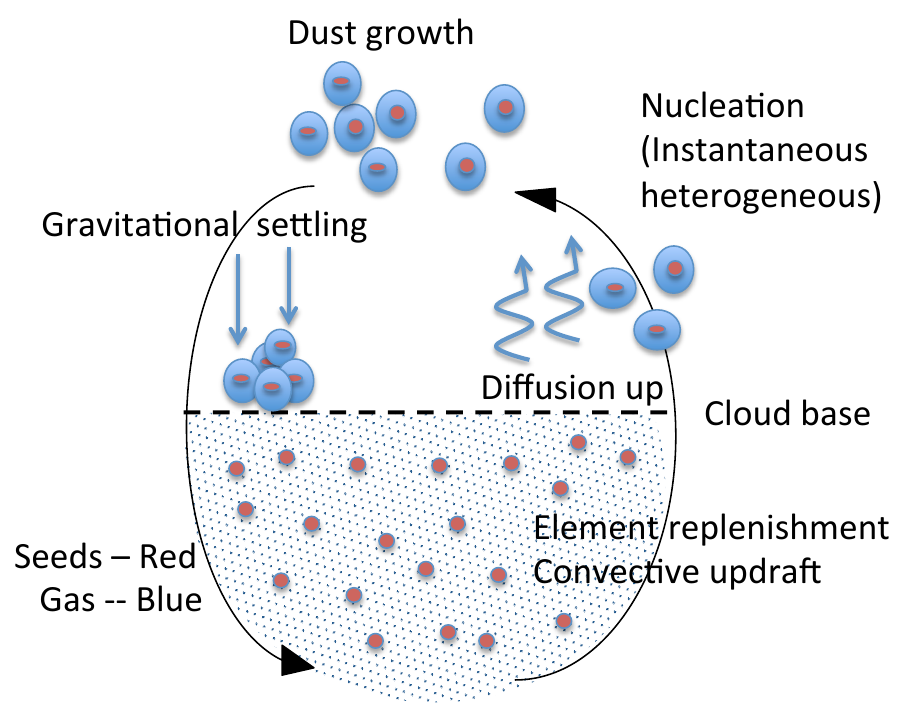}}
\caption{Schematic of dust growth, indicating instantaneous, heterogeneous condensation at a cloud base (dashed line) of ``blue" material on preexisting ``red" seeds of more refractory material, formed at lower altitudes. Monomers grow into aggregates by sticking, and all particles diffuse vertically and settle under gravity. Fluffy aggregates settle less readily than compact particles.}
\label{cartoon}
\vspace{-0.2in}
\end{figure}

We consider here a simplified scenario where the condensate seeds move upward, and the upper atmosphere is a stagnant layer sitting on top of a turbulent layer. The dominant modes of particle transport in the stagnant layer are Brownian diffusion and gravitational settling.  In the turbulent layer below, we adopt a cloud scenario where \textit{heterogeneous} nucleation (i.e., condensation upon pre-existing small particles of different composition) dominates under most realistic conditions \citep{Rossow1978,yair1995,movshovitz2008}. This assumes that there are cloud condensation nuclei (CCN), or ``seeds'', composed of more refractory material being diffused upward from below and that condensation upon these is rapid (see {\bf Figure \ref{cartoon}}, see also \citealt{Lee2016,Gao&Benneke2018,lee2018,helling2019} for further discussion of condensation efficiencies in exoplanet atmospheres). In this configuration, aggregates would form from the seeds coming from below via turbulence. Once these aggregates reach the stagnant boundary layer, they can continue to diffuse upward via Brownian motion.  

\begin{figure}[!ht]
\begin{center}
\includegraphics[width=0.99\textwidth]{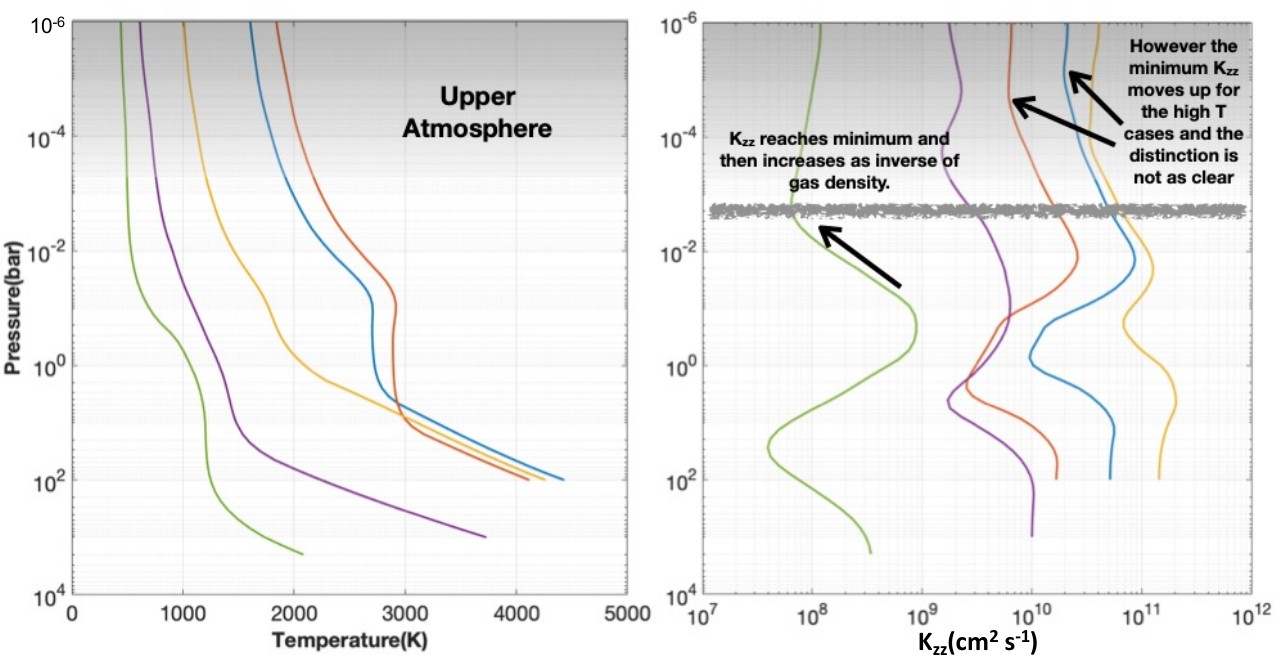}
\caption{Left: Temperature-pressure profiles for a suite of {irradiated hot atmospheres, from the \texttt{EGP+} model suite \citep{Fortney2008TwoClassesofIrradiatedAtmospheres} to demonstrate the parameter space. Right: Corresponding eddy diffusivities $K_\mathrm{zz}$ for the various atmospheres using the methods of \citet{SaumonMarley2008}. All models are solar metallicity}. The upper part of the atmosphere where $K_\mathrm{zz}$ reaches a minimum and then increases as the inverse of the gas density is highlighted {for the coolest atmosphere (green line)} as the grey chalked line. We assume aerosol transport via Brownian motion {above this region where $K_\mathrm{zz}$ transitions to nearly isothermal, the exact pressure of which depends on the specific temperature structure.}}
\label{atmos_kzz}
\end{center}
\vspace{-0.2in}
\end{figure}

In order to estimate the altitude of the stagnant layer in this simple scenario, we use vertical profiles of gas diffusivity $K_\mathrm{zz}$ {from \citet{SaumonMarley2008} for typical hot (1000 - 2500 K) atmospheres, which uses mixing length theory to derive $K_\mathrm{zz}$ at depth and then scales $K_\mathrm{zz}$ as the inverse of gas density above convective zones}. Turbulence causes high eddy diffusivity (as shown by large $K_\mathrm{zz}$) deeper in the atmosphere ({\bf Figure \ref{atmos_kzz}}). We expect there is some altitude where $K_\mathrm{zz}$ reaches a local minimum because the onset of the stagnant layer is dominated by Brownian diffusion instead of eddy-driven diffusion.

Brownian motion is governed by molecular collisions and the overall kinetic energy of particles, while settling is due to the gravitational force of the planet. The Brownian diffusivity for aerosol particles with radius ($r$) in a gaseous medium is defined as:

\begin{equation}
D_\mathrm{b}=kT {\beta \over 6\pi \eta r}
\label{Diff}
\end{equation}
where $T$ is the gas temperature, $k$ is the Boltzmann constant, $\beta$ is the Cunningham slip factor defined earlier in this section in Equation \ref{slipfactor}, and $\eta$ is the gas dynamical viscosity defined in Equation \ref{eq:eta}. As shown in the beginning of this section, the upper atmosphere is characterized by the kinetic/Epstein regime, where the Knudsen number ${K\!n}\gg1$, and the Cunningham factor becomes $\beta \sim K\!n$. Plugging in these values into the diffusivity equation (Eq. \ref{Diff}), we get an expression for particle diffusivity in the upper atmosphere via Brownian motion:

\begin{equation}
D_\mathrm{b}(kinetic)={kT \over 2 \pi r^{2} c \rho_\mathrm{g}}
\label{Diff_kinetic}
\end{equation}

We can compare the Brownian particle diffusivity ($D_\mathrm{b}$) for a stagnant layer with the the particle diffusivity derived from $K_\mathrm{zz}$ values for a turbulent atmosphere. The gas eddy diffusivity ($K_\mathrm{zz}$) is related to the particle diffusivity ($D^{p}_\mathrm{zz}$) by the Stokes number ($St$, defined below), where $D^{p}_\mathrm{zz}={K_\mathrm{zz} \over (1+St^{2})}$. If $K_\mathrm{zz}=l v_\mathrm{zz}$, where $l$ is the eddy length scale, and $v_\mathrm{zz}$ is the eddy velocity, then the Stokes number is defined as $St=t_\mathrm{s}\Omega$, where $\Omega={v_\mathrm{zz}\over l}$. The physics is that of a particle with a given response time responding to oscillatory forcing \citep{Volk1980,Cuzzi1993,Dubrulle1995,Carballido2011}. The expression for the particle diffusivity as a function of $K_\mathrm{zz}$ becomes:

\begin{equation}
D^{p}_\mathrm{zz}={K_\mathrm{zz} \over 1+St^{2} } = {K_\mathrm{zz} \over (1+ (t_\mathrm{s} {K_\mathrm{zz}\over H^{2} })^{2})} \,\,\,  \mbox{if we set $l=H$, and $v_\mathrm{zz}={K_\mathrm{zz}\over H}$}
\label{Dpzz}
\end{equation}
   
\begin{figure*}[h!]
\begin{center}
\includegraphics[angle=0,width=3.2in,height=2.7in]{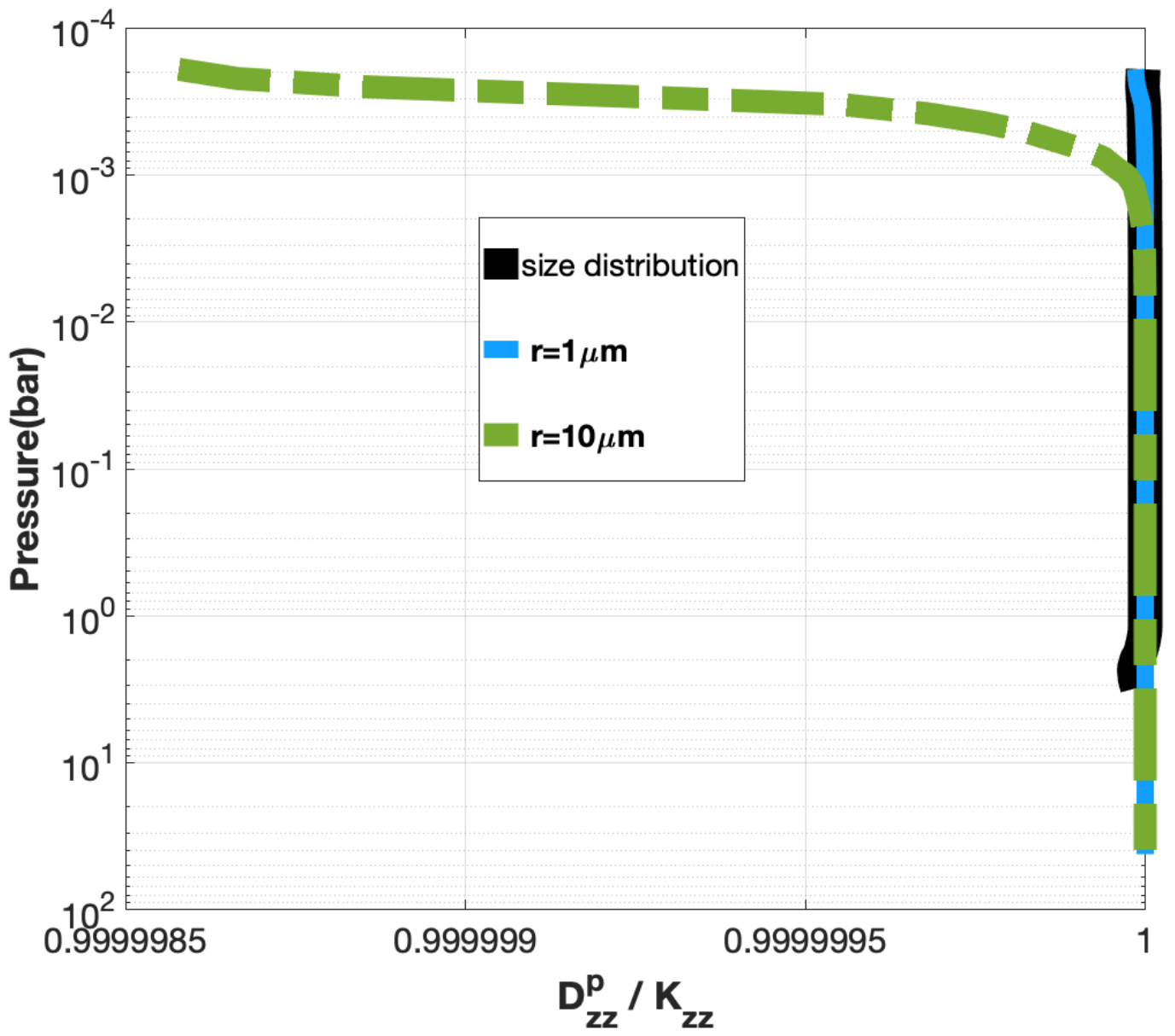}
\includegraphics[angle=0,width=3.2in,height=2.7in]{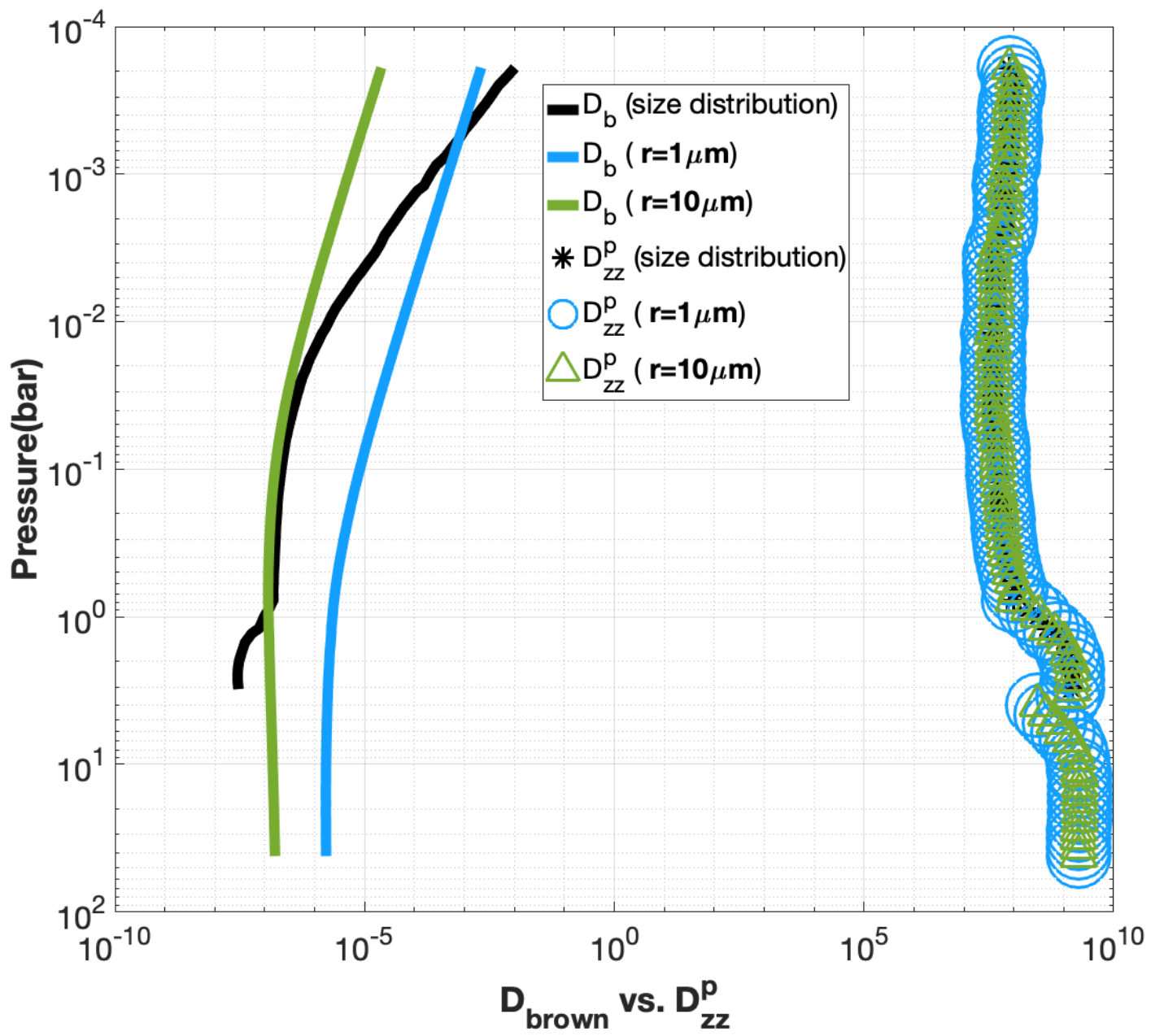}
\end{center}
\caption{Left: Ratio of particle diffusivity ($D^{p}_\mathrm{zz}$) to gas diffusivity ($K_\mathrm{zz}$) as a function of pressure for a sample atmosphere from Figure \ref{atmos_kzz}, with a particle size distribution as shown in black, and single sizes with radius $r=1\mu m$ in blue, and $r=10\mu m$ in green. The particle diffusivity ($D^{p}_\mathrm{zz}$) follows the gas diffusivity ($K_\mathrm{zz}$) throughout the atmosphere and diverges from unity in the upper atmosphere. Right: Brownian diffusion ($D_\mathrm{b}$) as a function of pressure, for a particle size distribution in black, and single sizes with radius $r=1\mu m$ in blue, and $r=10\mu m$ in green. This is compared to particle diffusion $D^{p}_\mathrm{zz}$ (shown in colored symbols). The different sizes of particles all have the approximately the same value, essentially following $K_\mathrm{zz}$. Brownian diffusion is much lower in magnitude than turbulent diffusion for expected values of $K_\mathrm{zz}$.}
\label{Db}
\end{figure*}

We calculate the particle eddy diffusivity ($D^{p}_\mathrm{zz}$) for a sample atmosphere with tabulated $K_\mathrm{zz}$ values using the Stokes number as defined in Equation \ref{Dpzz}. These calculations are done for different particle sizes and shown in {\bf Figure \ref{Db}}. We can see from {\bf Figure \ref{Db} (left)} that the ratio of Brownian particle diffusivity to $K_\mathrm{zz}$ is approximately unity throughout the atmosphere and only starts deviating negligibly from unity in the upper region, around millibar pressure levels{, for the largest particles}. The Brownian diffusivity ($D_\mathrm{b}$) is also plotted in {\bf Figure \ref{Db} (right)} for different particle sizes, and we can see that the magnitude of the Brownian motion diffusivity is much less than the $K_\mathrm{zz}$ diffusivity.

 {\citet{Woitke2020} similarly explores the transition between the region of vigorous mixing and Brownian motion, and finds that the onset of the Brownian motion-dominated regime occurs at 10$^{-6}$ bar rather than 10$^{-3}$ bar as we initially conservatively assume in Figure \ref{atmos_kzz}, at least for the coolest atmosphere. \citet{Woitke2020} presents models for hot Jupiters with T$_{\rm{eff}}$ = 2000 K and log(g) = 3, while our atmospheric models are for various temperatures from 1000 -- 2500 K and log(g) = 3 -- 4. \citet{Woitke2020} also specifically includes nucleation, growth, drift and diffusion for the generation and transport of particles, while we only consider diffusion and assume particles maintain their size once mixed upward from depth. These differences highlight that the exact point of transition to the homopause and the thermally dominated, rather than eddy diffusion dominated, regime depends on both the temperature structure of the object as well as the details of particle-particle interactions. However, as shown in Figure \ref{Db}, even our simpler models result in eddy-dominated regimes to at least 10$^{-5}$ bar in agreement with the more rigorous calculations performed in \citet{Woitke2020}.}
 
Given the efficiency of $K_\mathrm{zz}$, particle motion via turbulence and via gas upward motions dragging along particles clearly offers a viable mechanism by which particles can be transported to high altitudes. Even given the onset of the stagnant boundary layer, the magnitude of expected $K_\mathrm{zz}$ values is such that particles condensing from below can be readily dragged to regions of the atmosphere where they could generate flat transmission spectra. 

{Here, we made a few simplifying assumptions to our consideration of particle transport. We have inherently assumed radiatively passive particles; however, we also note that particle opacity can itself generate turbulence or convective motion if radiatively active particle feedback is large enough \citep[e.g.,][]{TanShowman2019,Lefevre2022,Lee2024}. Moreover, we are considering particle transport only in a 1-D sense, and do not account for 2-D effects such as global wind patterns or other non-diffusive large scale mixing, which could also be the dominant source of mixing in a real atmosphere.}

\begin{figure*}[h!]
\begin{center}
\includegraphics[width=0.62\textwidth]{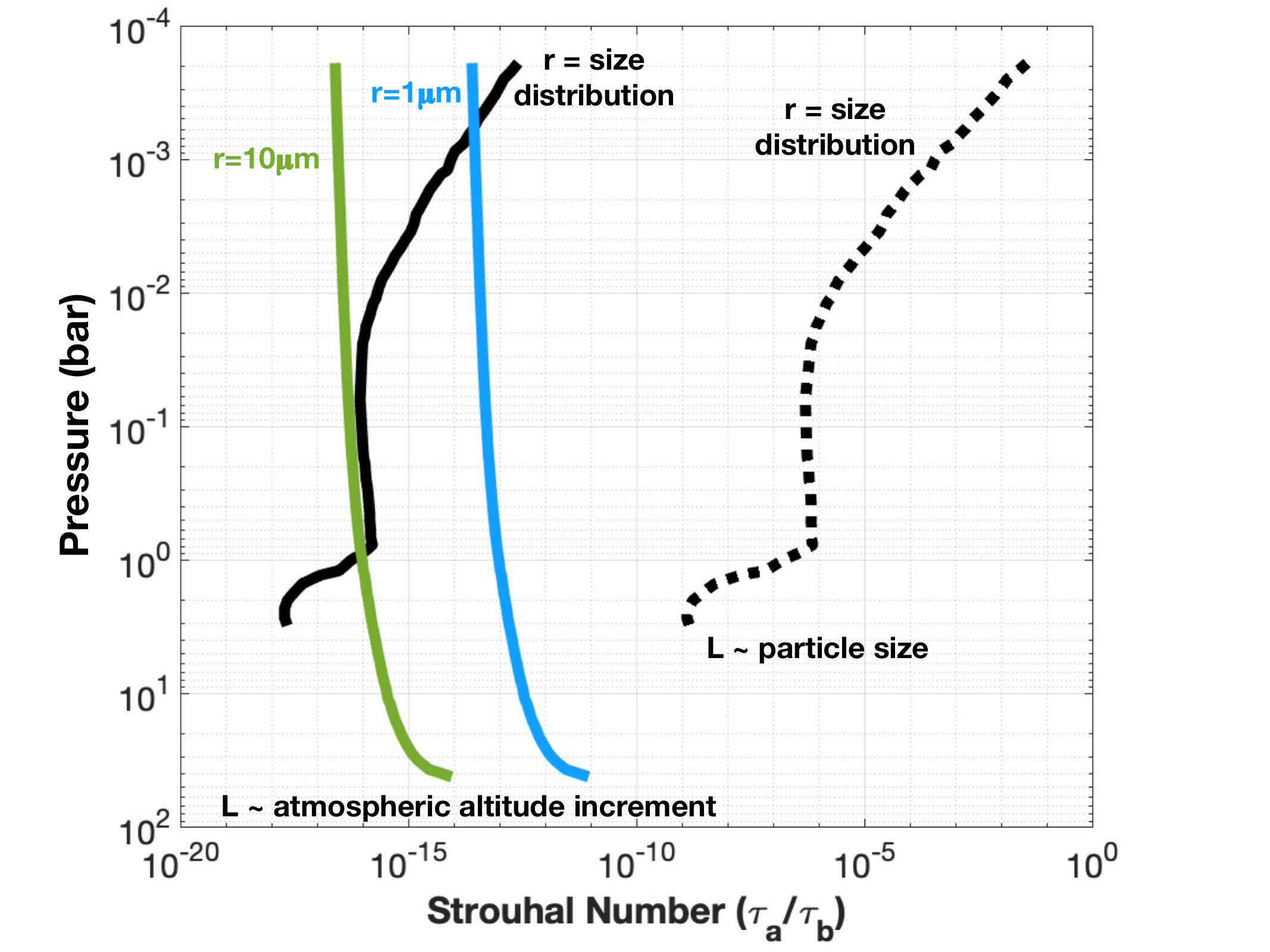}
\end{center}
\caption{Strouhal number for different particle sizes (particle size distribution in black, single sizes with $r=1 \mu m$ in blue and $r=10 \mu m$ in green). We can see that for all cases $Str\ll1$, which means Brownian motion is not enough to prevent gravitational settling. Also, the efficiency is dependent on the length scales we are comparing and increases for a shorter length scale ($L \sim r$) as shown in dotted black for the particle size distribution.}
\label{Str}
\end{figure*}

To further see that Brownian motion is not an effective means of transport against gravitational settling of aerosols, we compare their relative timescales to move a particle a given distance ($L$). The ratio of gravitational settling -- or the drift timescale ($t_\mathrm{d}$) -- to the Brownian motion timescale ($t_\mathrm{b}$) is called the Strouhal number ($Str={t_\mathrm{d} \over t_\mathrm{b}}$). The timescales are defined as the following:

\begin{equation}
t_\mathrm{d}={L \over v_\mathrm{t}} \,\,\,\,\,\,\, t_\mathrm{b}={L^{2} \over D_\mathrm{b}},  \mbox{ and}\,\, Str={D_\mathrm{b} \over Lv_\mathrm{t}}
\label{Strouhal}
\end{equation}
where $D_\mathrm{b}$ is the Brownian particle diffusion coefficient, and the terminal (or fall) velocity is $v_\mathrm{t}=gt_\mathrm{s}$. 
 If $Str\ll1$, then Brownian motion is negligible. We can see from {\bf Figure \ref{Str}} that Brownian motion is not an efficient mode of transport in the upper ({millibar -- 100 millibar pressure levels}) atmosphere -- at least for large particles (large meaning larger than molecular domain, see Section \ref{section:stop_time}). Indeed, it is differential velocity from Brownian motions that allows tiny particles to collide and grow. The transport efficiency also depends on the length of transport, as shown in {\bf Figure \ref{Str}}, where for shorter length scales, Brownian motion is more efficient than for longer distances. 
 

\newpage
\section{Summary}

In this work, we have summarized the generalized, overarching effects of non-homogenous, non-spherical {aerosol} particles in atmospheric models, as applied to exoplanet and brown dwarf atmospheres. Other works \citep{Adams2019,Ohno2020,samra2020,Samra2022} have delved further into microphysical interactions of non-spherical particles as the exoplanet literature joins the long-standing  solar system treatment of complex cloud {and haze} particle morphologies \citep[e.g.,][etc.]{Toon1980,zhang2013}. Here, we have focused on the primary optical and dynamical effects of such non-spherical, nonhomogeneous, porous particles, regardless of their formation mechanism. We have guided the reader through two major consequences of non-spherical, non-homogeneous particles.

First, for all particle shapes, large particles increase the extent of the wavelength independent regime. Upon discovering a flat transmission spectrum, one can consider how a more reasonably sized porous or fluffy aggregate particle could explain such an observation better than an overly large solid spherical particle. One would recall that:

\begin{itemize}
\setlength{\itemindent}{.5in}
\item Highly porous particles lofted to high altitudes can alter the onset of the wavelength independent regime, moving it to shorter wavelengths given a porous particle of the same mass as a solid particle -- Equation \ref{eq:phaseshift_porous} and Equation \ref{Opt_Phase_Rrho}.
\item Consequently, for a porous particle and solid particle \textit{of the same mass} both in the wavelength-independent regime, the porous particle will have a higher opacity, with the opacity increasing with increased porosity -- Equation \ref{eq_opacity} and Figure \ref{opacity}.
\item
The discrepancy between realistic aggregates modeled with the discrete dipole approximation (DDA) and homogeneous particles using Mie theory increases for porous/lacy aggregate structures (i.e., for fractal number  D $<$ 2) {\citep[e.g.,][]{Vahidinia2011}}. At higher porosities, the realistic DDA aggregate maintains the wavelength  independent regime out to longer wavelengths. This highlights the inherent degeneracy between number density, mass, porosity, and compactness of a given particle size and shape to explain observations -- Equations \ref{aggregate_mass_density}, \ref{aggregate_porosity}, and \ref{column_mass2}. {If using simple Mie-EMT, the particle mass might be overestimated, as compact particles require more particles overall to achieve the needed extinction.}
\end{itemize}

\noindent {Should the reader be interested in adding porous or aggregate opacity effects into existing frameworks, we suggest to alter existing Mie codes with EMT to account for porosity by following our Equation \ref{same_mass_r},  Equation \ref{same_mass}, and  Equation \ref{same_refract}  to adjust the effective particle radius, filling factor, and refractive index as necessary. The simple volume averaged EMT approach in Equation \ref{same_refract} is applicable for materials with refractive indices on order of unity and is valuable in capturing the impact of porosity. For materials with much higher refractive indices such as metals, this simple approach overestimates the higher refractive index component as shown in \citet{Cuzzi2014}, and the full Maxwell Garnett theory or other EMT variants can be used as referenced in section \ref{section:porANDcomp}.}

 Particle composition, and thus refractive index, can also compensate for porosity changes. To generate flat transmission spectra for exoplanets, highly porous or very lacy aggregates must have quite large real refractive indices. A reader may also recall recent observations of the sub-Neptune exoplanet GJ~1214b by JWST which invoked highly scattering (i.e., very high real refractive index) particles to match observations \citep{Gao2023}.  Neither those authors nor we here have delved into the compositions or mechanisms that would generate such high refractive index particles, but we encourage future studies along this line of inquiry.

Second, we have examined the dynamic regimes of aerosol particles throughout the atmosphere. In the case of a flat spectrum where a large mass of particles might seem unlikely due to any material quickly settling out under gravity, one would recall that:

\begin{itemize}
\setlength{\itemindent}{.5in}
\item Fluffy aggregate particles have shorter stopping times due to increased particle drag -- Equations \ref{stop_aggs31} and \ref{stop_aggs32} -- which results in such fluffy particles having longer residence lifetimes high in the atmosphere. 
\end{itemize}
Therefore, such particles are likely to contribute more to the opacity than faster-falling solid particles, and would also have a stronger effect on transmission spectra compared to compact, solid particles. 

We have also shown that 

\begin{itemize}
\setlength{\itemindent}{.5in}
\item Brownian motion is far less efficient than gravitational settling or particle drift for transporting these particles in the upper atmosphere over long distances -- Equation \ref{Diff_kinetic} and \ref{Dpzz} -- meaning Brownian motion does not serve as a way to remove these large fluffy aggregates from high altitudes.
\end{itemize}

We have shown that once particles exist at appropriately high altitudes, they have long residence times, but we have not explored in depth how such particles reach these locations in the first place. We suggest that either upward motion via gas diffusivity, $K_\mathrm{zz}$, of particles that are generated deeper in the atmosphere or \textit{in situ} production of particles via photochemical means are potential mechanisms, both of which are discussed in additional detail in various other studies \citep{Adams2019,Ohno2020,samra2020,Samra2022}. Future work focusing specifically on the role of upward mixing versus sedimentation efficiency for such fluffy aggregates is also planned (Moran et al. in preparation).

Combining both optical and dynamical effects, porous and/or aggregate particles can readily explain flat transmission spectra of exoplanets. Not only do such aggregate particles have stronger overall opacity for a given particle mass, but such particles are also likelier to persist over longer timescales at high altitudes. As recent work has already started exploring, we suggest that exoplanet atmospheric models continue to explore not only cloud and haze composition, atmospheric mixing, and particle size, but also particle morphology as a potential reason for muted and flat exoplanet transmission spectra.

\begin{acknowledgments}
The authors gratefully acknowledge T.D. Robinson for his help in obtaining old data thought to be lost to time.
\end{acknowledgments}

\bibliography{vahidiniamoran2024}{}

\begin{thebibliography}{}
\expandafter\ifx\csname natexlab\endcsname\relax\def\natexlab#1{#1}\fi
\providecommand{\url}[1]{\href{#1}{#1}}
\providecommand{\dodoi}[1]{doi:~\href{http://doi.org/#1}{\nolinkurl{#1}}}
\providecommand{\doeprint}[1]{\href{http://ascl.net/#1}{\nolinkurl{http://ascl.net/#1}}}
\providecommand{\doarXiv}[1]{\href{https://arxiv.org/abs/#1}{\nolinkurl{https://arxiv.org/abs/#1}}}

\bibitem[{{Ackerman} \& {Marley}(2001)}]{AM01}
{Ackerman}, A.~S., \& {Marley}, M.~S. 2001, \apj, 556, 872, \dodoi{10.1086/321540}

\bibitem[{{Adams} {et~al.}(2019){Adams}, {Gao}, {de Pater}, \& {Morley}}]{Adams2019}
{Adams}, D., {Gao}, P., {de Pater}, I., \& {Morley}, C.~V. 2019, \apj, 874, 61, \dodoi{10.3847/1538-4357/ab074c}

\bibitem[{{Bean} {et~al.}(2021){Bean}, {Raymond}, \& {Owen}}]{Bean2021}
{Bean}, J.~L., {Raymond}, S.~N., \& {Owen}, J.~E. 2021, Journal of Geophysical Research (Planets), 126, e06639, \dodoi{10.1029/2020JE006639}

\bibitem[{{Bean} {et~al.}(2011){Bean}, {D{\'e}sert}, {Kabath}, {Stalder}, {Seager}, {Miller-Ricci Kempton}, {Berta}, {Homeier}, {Walsh}, \& {Seifahrt}}]{bean2011}
{Bean}, J.~L., {D{\'e}sert}, J.-M., {Kabath}, P., {et~al.} 2011, \apj, 743, 92, \dodoi{10.1088/0004-637X/743/1/92}

\bibitem[{{Bohren} \& {Huffman}(1983)}]{BohrenHuffman1983}
{Bohren}, C.~F., \& {Huffman}, D.~R. 1983, {Absorption and scattering of light by small particles}

\bibitem[{{Brogi} {et~al.}(2016){Brogi}, {de Kok}, {Albrecht}, {Snellen}, {Birkby}, \& {Schwarz}}]{brogi2016}
{Brogi}, M., {de Kok}, R.~J., {Albrecht}, S., {et~al.} 2016, \apj, 817, 106, \dodoi{10.3847/0004-637X/817/2/106}

\bibitem[{{Burningham} {et~al.}(2017){Burningham}, {Marley}, {Line}, {Lupu}, {Visscher}, {Morley}, {Saumon}, \& {Freedman}}]{Burningham2017}
{Burningham}, B., {Marley}, M.~S., {Line}, M.~R., {et~al.} 2017, \mnras, 470, 1177, \dodoi{10.1093/mnras/stx1246}

\bibitem[{Cable {et~al.}(2012)Cable, Hörst, Hodyss, Beauchamp, Smith, \& Willis}]{cable2012}
Cable, M.~L., Hörst, S.~M., Hodyss, R., {et~al.} 2012, Chemical Reviews, 112, 1882, \dodoi{10.1021/cr200221x}

\bibitem[{{Carballido} {et~al.}(2011){Carballido}, {Bai}, \& {Cuzzi}}]{Carballido2011}
{Carballido}, A., {Bai}, X.-N., \& {Cuzzi}, J.~N. 2011, \mnras, 415, 93, \dodoi{10.1111/j.1365-2966.2011.18661.x}

\bibitem[{{Charnay} {et~al.}(2015){Charnay}, {Meadows}, {Misra}, {Leconte}, \& {Arney}}]{charnay2015}
{Charnay}, B., {Meadows}, V., {Misra}, A., {Leconte}, J., \& {Arney}, G. 2015, \apjl, 813, L1, \dodoi{10.1088/2041-8205/813/1/L1}

\bibitem[{{Cushing} {et~al.}(2010){Cushing}, {Saumon}, \& {Marley}}]{cushing2010}
{Cushing}, M.~C., {Saumon}, D., \& {Marley}, M.~S. 2010, \aj, 140, 1428, \dodoi{10.1088/0004-6256/140/5/1428}

\bibitem[{{Cuzzi}(1985)}]{Cuzzi1985}
{Cuzzi}, J.~N. 1985, \icarus, 63, 312, \dodoi{10.1016/0019-1035(85)90014-4}

\bibitem[{{Cuzzi} {et~al.}(1993){Cuzzi}, {Dobrovolskis}, \& {Champney}}]{Cuzzi1993}
{Cuzzi}, J.~N., {Dobrovolskis}, A.~R., \& {Champney}, J.~M. 1993, \icarus, 106, 102, \dodoi{10.1006/icar.1993.1161}

\bibitem[{{Cuzzi} {et~al.}(2014){Cuzzi}, {Estrada}, \& {Davis}}]{Cuzzi2014}
{Cuzzi}, J.~N., {Estrada}, P.~R., \& {Davis}, S.~S. 2014, \apjs, 210, 21, \dodoi{10.1088/0067-0049/210/2/21}

\bibitem[{{Cuzzi} \& {Pollack}(1978)}]{Cuzzi1978}
{Cuzzi}, J.~N., \& {Pollack}, J.~B. 1978, \icarus, 33, 233, \dodoi{10.1016/0019-1035(78)90145-8}

\bibitem[{{de Mooij} {et~al.}(2013){de Mooij}, {Brogi}, {de Kok}, {Snellen}, {Croll}, {Jayawardhana}, {Hoekstra}, {Otten}, {Bekkers}, {Haffert}, \& {van Houdt}}]{DeMooij2013}
{de Mooij}, E.~J.~W., {Brogi}, M., {de Kok}, R.~J., {et~al.} 2013, \apj, 771, 109, \dodoi{10.1088/0004-637X/771/2/109}

\bibitem[{{Demory} {et~al.}(2013){Demory}, {de Wit}, {Lewis}, {Fortney}, {Zsom}, {Seager}, {Knutson}, {Heng}, {Madhusudhan}, {Gillon}, {Barclay}, {Desert}, {Parmentier}, \& {Cowan}}]{Demory2013}
{Demory}, B.-O., {de Wit}, J., {Lewis}, N., {et~al.} 2013, \apjl, 776, L25, \dodoi{10.1088/2041-8205/776/2/L25}

\bibitem[{{Draine} \& {Flatau}(1994)}]{Draine1994}
{Draine}, B.~T., \& {Flatau}, P.~J. 1994, Journal of the Optical Society of America A, 11, 1491, \dodoi{10.1364/JOSAA.11.001491}

\bibitem[{{Draine} \& {Lee}(1984)}]{DraineLee1984}
{Draine}, B.~T., \& {Lee}, H.~M. 1984, \apj, 285, 89, \dodoi{10.1086/162480}

\bibitem[{{Dubrulle} {et~al.}(1995){Dubrulle}, {Morfill}, \& {Sterzik}}]{Dubrulle1995}
{Dubrulle}, B., {Morfill}, G., \& {Sterzik}, M. 1995, \icarus, 114, 237, \dodoi{10.1006/icar.1995.1058}

\bibitem[{{Evans} {et~al.}(2013){Evans}, {Pont}, {Sing}, {Aigrain}, {Barstow}, {D{\'e}sert}, {Gibson}, {Heng}, {Knutson}, \& {Lecavelier des Etangs}}]{Evans2013}
{Evans}, T.~M., {Pont}, F., {Sing}, D.~K., {et~al.} 2013, \apjl, 772, L16, \dodoi{10.1088/2041-8205/772/2/L16}

\bibitem[{{Feinstein} {et~al.}(2023){Feinstein}, {Radica}, {Welbanks}, {Murray}, {Ohno}, {Coulombe}, {Espinoza}, {Bean}, {Teske}, {Benneke}, {Line}, {Rustamkulov}, {Saba}, {Tsiaras}, {Barstow}, {Fortney}, {Gao}, {Knutson}, {MacDonald}, {Mikal-Evans}, {Rackham}, {Taylor}, {Parmentier}, {Batalha}, {Berta-Thompson}, {Carter}, {Changeat}, {dos Santos}, {Gibson}, {Goyal}, {Kreidberg}, {L{\'o}pez-Morales}, {Lothringer}, {Miguel}, {Molaverdikhani}, {Moran}, {Morello}, {Mukherjee}, {Sing}, {Stevenson}, {Wakeford}, {Ahrer}, {Alam}, {Alderson}, {Allen}, {Batalha}, {Bell}, {Blecic}, {Brande}, {Caceres}, {Casewell}, {Chubb}, {Crossfield}, {Crouzet}, {Cubillos}, {Decin}, {D{\'e}sert}, {Harrington}, {Heng}, {Henning}, {Iro}, {Kempton}, {Kendrew}, {Kirk}, {Krick}, {Lagage}, {Lendl}, {Mancini}, {Mansfield}, {May}, {Mayne}, {Nikolov}, {Palle}, {Petit dit de la Roche}, {Piaulet}, {Powell}, {Redfield}, {Rogers}, {Roman}, {Roy}, {Nixon}, {Schlawin}, {Tan}, {Tremblin}, {Turner}, {Venot}, {Waalkes}, {Wheatley}, \&
  {Zhang}}]{feinstein2023}
{Feinstein}, A.~D., {Radica}, M., {Welbanks}, L., {et~al.} 2023, \nat, 614, 670, \dodoi{10.1038/s41586-022-05674-1}

\bibitem[{{Fortney} {et~al.}(2008){Fortney}, {Lodders}, {Marley}, \& {Freedman}}]{Fortney2008TwoClassesofIrradiatedAtmospheres}
{Fortney}, J.~J., {Lodders}, K., {Marley}, M.~S., \& {Freedman}, R.~S. 2008, \apj, 678, 1419, \dodoi{10.1086/528370}

\bibitem[{{Fortney} {et~al.}(2013){Fortney}, {Mordasini}, {Nettelmann}, {Kempton}, {Greene}, \& {Zahnle}}]{Fortney2013}
{Fortney}, J.~J., {Mordasini}, C., {Nettelmann}, N., {et~al.} 2013, \apj, 775, 80, \dodoi{10.1088/0004-637X/775/1/80}

\bibitem[{{Gao} \& {Benneke}(2018)}]{Gao&Benneke2018}
{Gao}, P., \& {Benneke}, B. 2018, \apj, 863, 165, \dodoi{10.3847/1538-4357/aad461}

\bibitem[{{Gao} \& {Powell}(2021)}]{Gao&Powell2021}
{Gao}, P., \& {Powell}, D. 2021, \apjl, 918, L7, \dodoi{10.3847/2041-8213/ac139f}

\bibitem[{{Gao} {et~al.}(2021){Gao}, {Wakeford}, {Moran}, \& {Parmentier}}]{gao2021}
{Gao}, P., {Wakeford}, H.~R., {Moran}, S.~E., \& {Parmentier}, V. 2021, Journal of Geophysical Research (Planets), 126, e06655, \dodoi{10.1029/2020JE006655}

\bibitem[{{Gao} {et~al.}(2020){Gao}, {Thorngren}, {Lee}, {Fortney}, {Morley}, {Wakeford}, {Powell}, {Stevenson}, \& {Zhang}}]{gao2020}
{Gao}, P., {Thorngren}, D.~P., {Lee}, E. K.~H., {et~al.} 2020, Nature Astronomy, 4, 951, \dodoi{10.1038/s41550-020-1114-3}

\bibitem[{{Gao} {et~al.}(2023){Gao}, {Piette}, {Steinrueck}, {Nixon}, {Zhang}, {Kempton}, {Bean}, {Rauscher}, {Parmentier}, {Batalha}, {Savel}, {Arnold}, {Roman}, {Malsky}, \& {Taylor}}]{Gao2023}
{Gao}, P., {Piette}, A. A.~A., {Steinrueck}, M.~E., {et~al.} 2023, \apj, 951, 96, \dodoi{10.3847/1538-4357/acd16f}

\bibitem[{{Garnett}(1904)}]{Garnett1904}
{Garnett}, J.~C.~M. 1904, Philosophical Transactions of the Royal Society of London Series A, 203, 385, \dodoi{10.1098/rsta.1904.0024}

\bibitem[{{Gibson} {et~al.}(2012){Gibson}, {Aigrain}, {Pont}, {Sing}, {D{\'e}sert}, {Evans}, {Henry}, {Husnoo}, \& {Knutson}}]{Gibson2012}
{Gibson}, N.~P., {Aigrain}, S., {Pont}, F., {et~al.} 2012, \mnras, 422, 753, \dodoi{10.1111/j.1365-2966.2012.20655.x}

\bibitem[{{Helling}(2019)}]{helling2019}
{Helling}, C. 2019, Annual Review of Earth and Planetary Sciences, 47, 583, \dodoi{10.1146/annurev-earth-053018-060401}

\bibitem[{{Helling} \& {Woitke}(2006)}]{helling2006}
{Helling}, C., \& {Woitke}, P. 2006, \aap, 455, 325, \dodoi{10.1051/0004-6361:20054598}

\bibitem[{{Helling} {et~al.}(2008){Helling}, {Woitke}, \& {Thi}}]{Helling2008}
{Helling}, C., {Woitke}, P., \& {Thi}, W.~F. 2008, \aap, 485, 547, \dodoi{10.1051/0004-6361:20078220}

\bibitem[{{Huitson} {et~al.}(2012){Huitson}, {Sing}, {Vidal-Madjar}, {Ballester}, {Lecavelier des Etangs}, {D{\'e}sert}, \& {Pont}}]{Huitson2012}
{Huitson}, C.~M., {Sing}, D.~K., {Vidal-Madjar}, A., {et~al.} 2012, \mnras, 422, 2477, \dodoi{10.1111/j.1365-2966.2012.20805.x}

\bibitem[{{Ingraham} {et~al.}(2014){Ingraham}, {Marley}, {Saumon}, {Marois}, {Macintosh}, {Barman}, {Bauman}, {Burrows}, {Chilcote}, {De Rosa}, {Dillon}, {Doyon}, {Dunn}, {Erikson}, {Fitzgerald}, {Gavel}, {Goodsell}, {Graham}, {Hartung}, {Hibon}, {Kalas}, {Konopacky}, {Larkin}, {Maire}, {Marchis}, {McBride}, {Millar-Blanchaer}, {Morzinski}, {Norton}, {Oppenheimer}, {Palmer}, {Patience}, {Perrin}, {Poyneer}, {Pueyo}, {Rantakyr{\"o}}, {Sadakuni}, {Saddlemyer}, {Savransky}, {Soummer}, {Sivaramakrishnan}, {Song}, {Thomas}, {Wallace}, {Wiktorowicz}, \& {Wolff}}]{Ingraham2014}
{Ingraham}, P., {Marley}, M.~S., {Saumon}, D., {et~al.} 2014, \apjl, 794, L15, \dodoi{10.1088/2041-8205/794/1/L15}

\bibitem[{{Kataoka} {et~al.}(2014){Kataoka}, {Okuzumi}, {Tanaka}, \& {Nomura}}]{Kataoka2014}
{Kataoka}, A., {Okuzumi}, S., {Tanaka}, H., \& {Nomura}, H. 2014, \aap, 568, A42, \dodoi{10.1051/0004-6361/201323199}

\bibitem[{{Kempton} {et~al.}(2023){Kempton}, {Zhang}, {Bean}, {Steinrueck}, {Piette}, {Parmentier}, {Malsky}, {Roman}, {Rauscher}, {Gao}, {Bell}, {Xue}, {Taylor}, {Savel}, {Arnold}, {Nixon}, {Stevenson}, {Mansfield}, {Kendrew}, {Zieba}, {Ducrot}, {Dyrek}, {Lagage}, {Stassun}, {Henry}, {Barman}, {Lupu}, {Malik}, {Kataria}, {Ih}, {Fu}, {Welbanks}, \& {McGill}}]{kempton2023}
{Kempton}, E. M.~R., {Zhang}, M., {Bean}, J.~L., {et~al.} 2023, \nat, 620, 67, \dodoi{10.1038/s41586-023-06159-5}

\bibitem[{{Kimura} {et~al.}(2006){Kimura}, {Kolokolova}, \& {Mann}}]{kimura2006}
{Kimura}, H., {Kolokolova}, L., \& {Mann}, I. 2006, \aap, 449, 1243, \dodoi{10.1051/0004-6361:20041783}

\bibitem[{{Kitzmann} \& {Heng}(2018)}]{KitzmannHeng2018}
{Kitzmann}, D., \& {Heng}, K. 2018, \mnras, 475, 94, \dodoi{10.1093/mnras/stx3141}

\bibitem[{{Knutson} {et~al.}(2014){Knutson}, {Benneke}, {Deming}, \& {Homeier}}]{Knutson2014}
{Knutson}, H.~A., {Benneke}, B., {Deming}, D., \& {Homeier}, D. 2014, \nat, 505, 66, \dodoi{10.1038/nature12887}

\bibitem[{{Kopparla} {et~al.}(2016){Kopparla}, {Natraj}, {Zhang}, {Swain}, {Wiktorowicz}, \& {Yung}}]{Kopparla2016}
{Kopparla}, P., {Natraj}, V., {Zhang}, X., {et~al.} 2016, \apj, 817, 32, \dodoi{10.3847/0004-637X/817/1/32}

\bibitem[{{Kreidberg} {et~al.}(2014){Kreidberg}, {Bean}, {D{\'e}sert}, {Benneke}, {Deming}, {Stevenson}, {Seager}, {Berta-Thompson}, {Seifahrt}, \& {Homeier}}]{Kreidberg2014}
{Kreidberg}, L., {Bean}, J.~L., {D{\'e}sert}, J.-M., {et~al.} 2014, \nat, 505, 69, \dodoi{10.1038/nature12888}

\bibitem[{{Lavvas} {et~al.}(2019){Lavvas}, {Koskinen}, {Steinrueck}, {Garc{\'\i}a Mu{\~n}oz}, \& {Showman}}]{Lavvas2019}
{Lavvas}, P., {Koskinen}, T., {Steinrueck}, M.~E., {Garc{\'\i}a Mu{\~n}oz}, A., \& {Showman}, A.~P. 2019, \apj, 878, 118, \dodoi{10.3847/1538-4357/ab204e}

\bibitem[{{Lavvas} {et~al.}(2011){Lavvas}, {Sander}, {Kraft}, \& {Imanaka}}]{Lavvas2011}
{Lavvas}, P., {Sander}, M., {Kraft}, M., \& {Imanaka}, H. 2011, \apj, 728, 80, \dodoi{10.1088/0004-637X/728/2/80}

\bibitem[{{Lee} {et~al.}(2016){Lee}, {Dobbs-Dixon}, {Helling}, {Bognar}, \& {Woitke}}]{Lee2016}
{Lee}, E., {Dobbs-Dixon}, I., {Helling}, C., {Bognar}, K., \& {Woitke}, P. 2016, \aap, 594, A48, \dodoi{10.1051/0004-6361/201628606}

\bibitem[{{Lee} {et~al.}(2018){Lee}, {Blecic}, \& {Helling}}]{lee2018}
{Lee}, E.~K.~H., {Blecic}, J., \& {Helling}, C. 2018, \aap, 614, A126, \dodoi{10.1051/0004-6361/201731977}

\bibitem[{{Lee} {et~al.}(2024){Lee}, {Tan}, \& {Tsai}}]{Lee2024}
{Lee}, E. K.~H., {Tan}, X., \& {Tsai}, S.-M. 2024, \mnras, 529, 2686, \dodoi{10.1093/mnras/stae537}

\bibitem[{{Lef{\`e}vre} {et~al.}(2022){Lef{\`e}vre}, {Tan}, {Lee}, \& {Pierrehumbert}}]{Lefevre2022}
{Lef{\`e}vre}, M., {Tan}, X., {Lee}, E. K.~H., \& {Pierrehumbert}, R.~T. 2022, \apj, 929, 153, \dodoi{10.3847/1538-4357/ac5e2d}

\bibitem[{Lodders {et~al.}(2009)Lodders, Palme, \& Gail}]{lodders2009landolt}
Lodders, K., Palme, H., \& Gail, H. 2009, JE Tr{\"u}mper, 4, 44

\bibitem[{{Lodge} {et~al.}(2024){Lodge}, {Wakeford}, \& {Leinhardt}}]{lodge2024}
{Lodge}, M.~G., {Wakeford}, H.~R., \& {Leinhardt}, Z.~M. 2024, \mnras, 527, 11113, \dodoi{10.1093/mnras/stad3743}

\bibitem[{{Marley} {et~al.}(2013){Marley}, {Ackerman}, {Cuzzi}, \& {Kitzmann}}]{marley2013}
{Marley}, M.~S., {Ackerman}, A.~S., {Cuzzi}, J.~N., \& {Kitzmann}, D. 2013, in Comparative Climatology of Terrestrial Planets, ed. S.~J. {Mackwell}, A.~A. {Simon-Miller}, J.~W. {Harder}, \& M.~A. {Bullock}, 367--392, \dodoi{10.2458/azu_uapress_9780816530595-ch015}

\bibitem[{{Marley} {et~al.}(1999){Marley}, {Gelino}, {Stephens}, {Lunine}, \& {Freedman}}]{Marley1999}
{Marley}, M.~S., {Gelino}, C., {Stephens}, D., {Lunine}, J.~I., \& {Freedman}, R. 1999, \apj, 513, 879, \dodoi{10.1086/306881}

\bibitem[{{Marley} \& {Robinson}(2015)}]{marleyrobinson2015}
{Marley}, M.~S., \& {Robinson}, T.~D. 2015, \araa, 53, 279, \dodoi{10.1146/annurev-astro-082214-122522}

\bibitem[{{Marley} {et~al.}(2021){Marley}, {Saumon}, {Visscher}, {Lupu}, {Freedman}, {Morley}, {Fortney}, {Seay}, {Smith}, {Teal}, \& {Wang}}]{Marley2021}
{Marley}, M.~S., {Saumon}, D., {Visscher}, C., {et~al.} 2021, \apj, 920, 85, \dodoi{10.3847/1538-4357/ac141d}

\bibitem[{{Miles} {et~al.}(2023){Miles}, {Biller}, {Patapis}, {Worthen}, {Rickman}, {Hoch}, {Skemer}, {Perrin}, {Whiteford}, {Chen}, {Sargent}, {Mukherjee}, {Morley}, {Moran}, {Bonnefoy}, {Petrus}, {Carter}, {Choquet}, {Hinkley}, {Ward-Duong}, {Leisenring}, {Millar-Blanchaer}, {Pueyo}, {Ray}, {Sallum}, {Stapelfeldt}, {Stone}, {Wang}, {Absil}, {Balmer}, {Boccaletti}, {Bonavita}, {Booth}, {Bowler}, {Chauvin}, {Christiaens}, {Currie}, {Danielski}, {Fortney}, {Girard}, {Grady}, {Greenbaum}, {Henning}, {Hines}, {Janson}, {Kalas}, {Kammerer}, {Kennedy}, {Kenworthy}, {Kervella}, {Lagage}, {Lew}, {Liu}, {Macintosh}, {Marino}, {Marley}, {Marois}, {Matthews}, {Matthews}, {Mawet}, {McElwain}, {Metchev}, {Meyer}, {Molliere}, {Pantin}, {Quirrenbach}, {Rebollido}, {Ren}, {Schneider}, {Vasist}, {Wyatt}, {Zhou}, {Briesemeister}, {Bryan}, {Calissendorff}, {Cantalloube}, {Cugno}, {De Furio}, {Dupuy}, {Factor}, {Faherty}, {Fitzgerald}, {Franson}, {Gonzales}, {Hood}, {Howe}, {Kraus}, {Kuzuhara}, {Lagrange}, {Lawson}, {Lazzoni},
  {Liu}, {Llop-Sayson}, {Lloyd}, {Martinez}, {Mazoyer}, {Quanz}, {Redai}, {Samland}, {Schlieder}, {Tamura}, {Tan}, {Uyama}, {Vigan}, {Vos}, {Wagner}, {Wolff}, {Ygouf}, {Zhang}, {Zhang}, \& {Zhang}}]{miles2023}
{Miles}, B.~E., {Biller}, B.~A., {Patapis}, P., {et~al.} 2023, \apjl, 946, L6, \dodoi{10.3847/2041-8213/acb04a}

\bibitem[{{Miller-Ricci Kempton} {et~al.}(2012{\natexlab{a}}){Miller-Ricci Kempton}, {Zahnle}, \& {Fortney}}]{mr-kempton2012}
{Miller-Ricci Kempton}, E., {Zahnle}, K., \& {Fortney}, J.~J. 2012{\natexlab{a}}, \apj, 745, 3, \dodoi{10.1088/0004-637X/745/1/3}

\bibitem[{{Miller-Ricci Kempton} {et~al.}(2012{\natexlab{b}}){Miller-Ricci Kempton}, {Zahnle}, \& {Fortney}}]{MillerRicciKempton2012}
---. 2012{\natexlab{b}}, \apj, 745, 3, \dodoi{10.1088/0004-637X/745/1/3}

\bibitem[{{Min} {et~al.}(2006){Min}, {Dominik}, {Hovenier}, {de Koter}, \& {Waters}}]{min2006}
{Min}, M., {Dominik}, C., {Hovenier}, J.~W., {de Koter}, A., \& {Waters}, L.~B.~F.~M. 2006, \aap, 445, 1005, \dodoi{10.1051/0004-6361:20053212}

\bibitem[{{Min} {et~al.}(2003){Min}, {Hovenier}, \& {de Koter}}]{Min2003}
{Min}, M., {Hovenier}, J.~W., \& {de Koter}, A. 2003, \aap, 404, 35, \dodoi{10.1051/0004-6361:20030456}

\bibitem[{{Min} {et~al.}(2005){Min}, {Hovenier}, \& {de Koter}}]{Min2005}
---. 2005, \aap, 432, 909, \dodoi{10.1051/0004-6361:20041920}

\bibitem[{{Min} {et~al.}(2020){Min}, {Ormel}, {Chubb}, {Helling}, \& {Kawashima}}]{Min2020}
{Min}, M., {Ormel}, C.~W., {Chubb}, K., {Helling}, C., \& {Kawashima}, Y. 2020, \aap, 642, A28, \dodoi{10.1051/0004-6361/201937377}

\bibitem[{{Min} {et~al.}(2016){Min}, {Rab}, {Woitke}, {Dominik}, \& {M{\'e}nard}}]{min2016}
{Min}, M., {Rab}, C., {Woitke}, P., {Dominik}, C., \& {M{\'e}nard}, F. 2016, \aap, 585, A13, \dodoi{10.1051/0004-6361/201526048}

\bibitem[{{Molli{\`e}re} {et~al.}(2020){Molli{\`e}re}, {Stolker}, {Lacour}, {Otten}, {Shangguan}, {Charnay}, {Molyarova}, {Nowak}, {Henning}, {Marleau}, {Semenov}, {van Dishoeck}, {Eisenhauer}, {Garcia}, {Garcia Lopez}, {Girard}, {Greenbaum}, {Hinkley}, {Kervella}, {Kreidberg}, {Maire}, {Nasedkin}, {Pueyo}, {Snellen}, {Vigan}, {Wang}, {de Zeeuw}, \& {Zurlo}}]{Molliere2020}
{Molli{\`e}re}, P., {Stolker}, T., {Lacour}, S., {et~al.} 2020, \aap, 640, A131, \dodoi{10.1051/0004-6361/202038325}

\bibitem[{{Morley} {et~al.}(2013){Morley}, {Fortney}, {Kempton}, {Marley}, {Visscher}, \& {Zahnle}}]{morley2013}
{Morley}, C.~V., {Fortney}, J.~J., {Kempton}, E. M.~R., {et~al.} 2013, \apj, 775, 33, \dodoi{10.1088/0004-637X/775/1/33}

\bibitem[{{Morley} {et~al.}(2018){Morley}, {Skemer}, {Allers}, {Marley}, {Faherty}, {Visscher}, {Beiler}, {Miles}, {Lupu}, {Freedman}, {Fortney}, {Geballe}, \& {Bjoraker}}]{morley2018}
{Morley}, C.~V., {Skemer}, A.~J., {Allers}, K.~N., {et~al.} 2018, \apj, 858, 97, \dodoi{10.3847/1538-4357/aabe8b}

\bibitem[{{Moses}(2014)}]{moses2014}
{Moses}, J.~I. 2014, Philosophical Transactions of the Royal Society of London Series A, 372, 20130073, \dodoi{10.1098/rsta.2013.0073}

\bibitem[{{Moses} {et~al.}(2013){Moses}, {Line}, {Visscher}, {Richardson}, {Nettelmann}, {Fortney}, {Barman}, {Stevenson}, \& {Madhusudhan}}]{Moses2013}
{Moses}, J.~I., {Line}, M.~R., {Visscher}, C., {et~al.} 2013, \apj, 777, 34, \dodoi{10.1088/0004-637X/777/1/34}

\bibitem[{{Movshovitz} \& {Podolak}(2008)}]{movshovitz2008}
{Movshovitz}, N., \& {Podolak}, M. 2008, \icarus, 194, 368, \dodoi{10.1016/j.icarus.2007.09.018}

\bibitem[{{Nasedkin} {et~al.}(2024){Nasedkin}, {Molli{\`e}re}, \& {Blain}}]{Nasedkin2024}
{Nasedkin}, E., {Molli{\`e}re}, P., \& {Blain}, D. 2024, The Journal of Open Source Software, 9, 5875, \dodoi{10.21105/joss.05875}

\bibitem[{{Ohno} \& {Kawashima}(2020)}]{ohno&kawashima2020}
{Ohno}, K., \& {Kawashima}, Y. 2020, \apjl, 895, L47, \dodoi{10.3847/2041-8213/ab93d7}

\bibitem[{{Ohno} \& {Okuzumi}(2018)}]{Ohno2018}
{Ohno}, K., \& {Okuzumi}, S. 2018, \apj, 859, 34, \dodoi{10.3847/1538-4357/aabee3}

\bibitem[{{Ohno} {et~al.}(2020){Ohno}, {Okuzumi}, \& {Tazaki}}]{Ohno2020}
{Ohno}, K., {Okuzumi}, S., \& {Tazaki}, R. 2020, \apj, 891, 131, \dodoi{10.3847/1538-4357/ab44bd}

\bibitem[{{Okuzumi} {et~al.}(2009){Okuzumi}, {Tanaka}, \& {Sakagami}}]{Okuzumi2009}
{Okuzumi}, S., {Tanaka}, H., \& {Sakagami}, M.-a. 2009, \apj, 707, 1247, \dodoi{10.1088/0004-637X/707/2/1247}

\bibitem[{{Ossenkopf}(1991)}]{ossenkopf1991}
{Ossenkopf}, V. 1991, \aap, 251, 210

\bibitem[{{Pinhas} \& {Madhusudhan}(2017)}]{Pinhas2017}
{Pinhas}, A., \& {Madhusudhan}, N. 2017, \mnras, 471, 4355, \dodoi{10.1093/mnras/stx1849}

\bibitem[{{Pollack} \& {Cuzzi}(1980)}]{Pollack1980}
{Pollack}, J.~B., \& {Cuzzi}, J.~N. 1980, Journal of the Atmospheric Sciences, 37, 868, \dodoi{10.1175/1520-0469(1980)037<0868:SBNPOS>2.0.CO;2}

\bibitem[{{Pollack} {et~al.}(1994){Pollack}, {Hollenbach}, {Beckwith}, {Simonelli}, {Roush}, \& {Fong}}]{Pollack1994}
{Pollack}, J.~B., {Hollenbach}, D., {Beckwith}, S., {et~al.} 1994, \apj, 421, 615, \dodoi{10.1086/173677}

\bibitem[{{Pont} {et~al.}(2013){Pont}, {Sing}, {Gibson}, {Aigrain}, {Henry}, \& {Husnoo}}]{Pont2013}
{Pont}, F., {Sing}, D.~K., {Gibson}, N.~P., {et~al.} 2013, \mnras, 432, 2917, \dodoi{10.1093/mnras/stt651}

\bibitem[{{Rossow}(1978)}]{Rossow1978}
{Rossow}, W.~B. 1978, \icarus, 36, 1, \dodoi{10.1016/0019-1035(78)90072-6}

\bibitem[{{Samra} {et~al.}(2022){Samra}, {Helling}, \& {Birnstiel}}]{Samra2022}
{Samra}, D., {Helling}, C., \& {Birnstiel}, T. 2022, \aap, 663, A47, \dodoi{10.1051/0004-6361/202142651}

\bibitem[{{Samra} {et~al.}(2020){Samra}, {Helling}, \& {Min}}]{samra2020}
{Samra}, D., {Helling}, C., \& {Min}, M. 2020, \aap, 639, A107, \dodoi{10.1051/0004-6361/202037553}

\bibitem[{{Saumon} \& {Marley}(2008)}]{SaumonMarley2008}
{Saumon}, D., \& {Marley}, M.~S. 2008, \apj, 689, 1327, \dodoi{10.1086/592734}

\bibitem[{{Seager} \& {Deming}(2010)}]{seager2010}
{Seager}, S., \& {Deming}, D. 2010, \araa, 48, 631, \dodoi{10.1146/annurev-astro-081309-130837}

\bibitem[{{Sing} {et~al.}(2011){Sing}, {Pont}, {Aigrain}, {Charbonneau}, {D{\'e}sert}, {Gibson}, {Gilliland}, {Hayek}, {Henry}, {Knutson}, {Lecavelier Des Etangs}, {Mazeh}, \& {Shporer}}]{sing2011}
{Sing}, D.~K., {Pont}, F., {Aigrain}, S., {et~al.} 2011, \mnras, 416, 1443, \dodoi{10.1111/j.1365-2966.2011.19142.x}

\bibitem[{{Sing} {et~al.}(2016){Sing}, {Fortney}, {Nikolov}, {Wakeford}, {Kataria}, {Evans}, {Aigrain}, {Ballester}, {Burrows}, {Deming}, {D{\'e}sert}, {Gibson}, {Henry}, {Huitson}, {Knutson}, {Lecavelier Des Etangs}, {Pont}, {Showman}, {Vidal-Madjar}, {Williamson}, \& {Wilson}}]{sing2016}
{Sing}, D.~K., {Fortney}, J.~J., {Nikolov}, N., {et~al.} 2016, \nat, 529, 59, \dodoi{10.1038/nature16068}

\bibitem[{{Steinrueck} {et~al.}(2021){Steinrueck}, {Showman}, {Lavvas}, {Koskinen}, {Tan}, \& {Zhang}}]{steinrueck2021}
{Steinrueck}, M.~E., {Showman}, A.~P., {Lavvas}, P., {et~al.} 2021, \mnras, 504, 2783, \dodoi{10.1093/mnras/stab1053}

\bibitem[{{Stognienko} {et~al.}(1995){Stognienko}, {Henning}, \& {Ossenkopf}}]{Stognienko1995}
{Stognienko}, R., {Henning}, T., \& {Ossenkopf}, V. 1995, \aap, 296, 797

\bibitem[{{Tan} \& {Showman}(2019)}]{TanShowman2019}
{Tan}, X., \& {Showman}, A.~P. 2019, \apj, 874, 111, \dodoi{10.3847/1538-4357/ab0c07}

\bibitem[{{Tazaki}(2021)}]{tazaki2021}
{Tazaki}, R. 2021, \mnras, 504, 2811, \dodoi{10.1093/mnras/stab1069}

\bibitem[{{Tazaki} \& {Tanaka}(2018)}]{tazaki2018}
{Tazaki}, R., \& {Tanaka}, H. 2018, \apj, 860, 79, \dodoi{10.3847/1538-4357/aac32d}

\bibitem[{{Tomasko} {et~al.}(2008){Tomasko}, {Doose}, {Engel}, {Dafoe}, {West}, {Lemmon}, {Karkoschka}, \& {See}}]{Tomasko2008}
{Tomasko}, M.~G., {Doose}, L., {Engel}, S., {et~al.} 2008, \planss, 56, 669, \dodoi{10.1016/j.pss.2007.11.019}

\bibitem[{{Toon} {et~al.}(1980){Toon}, {Turco}, \& {Pollack}}]{Toon1980}
{Toon}, O.~B., {Turco}, R.~P., \& {Pollack}, J.~B. 1980, \icarus, 43, 260, \dodoi{10.1016/0019-1035(80)90173-6}

\bibitem[{{Vahidinia} {et~al.}(2014){Vahidinia}, {Cuzzi}, {Marley}, \& {Fortney}}]{Vahidinia2014}
{Vahidinia}, S., {Cuzzi}, J.~N., {Marley}, M., \& {Fortney}, J. 2014, \apjl, 789, L11, \dodoi{10.1088/2041-8205/789/1/L11}

\bibitem[{{Vahidinia} {et~al.}(2011){Vahidinia}, {Cuzzi}, {Hedman}, {Draine}, {Clark}, {Roush}, {Filacchione}, {Nicholson}, {Brown}, {Buratti}, \& {Sotin}}]{Vahidinia2011}
{Vahidinia}, S., {Cuzzi}, J.~N., {Hedman}, M., {et~al.} 2011, \icarus, 215, 682, \dodoi{10.1016/j.icarus.2011.04.011}

\bibitem[{Van~de Hulst(1981)}]{hulst1981}
Van~de Hulst, H.~C. 1981, Light Scattering by Small Particles (Courier Corporation)

\bibitem[{{Voelk} {et~al.}(1980){Voelk}, {Jones}, {Morfill}, \& {Roeser}}]{Volk1980}
{Voelk}, H.~J., {Jones}, F.~C., {Morfill}, G.~E., \& {Roeser}, S. 1980, \aap, 85, 316

\bibitem[{{Volten} {et~al.}(2007){Volten}, {Mu{\~n}oz}, {Hovenier}, {Rietmeijer}, {Nuth}, {Waters}, \& {van der Zande}}]{Volten2007}
{Volten}, H., {Mu{\~n}oz}, O., {Hovenier}, J.~W., {et~al.} 2007, \aap, 470, 377, \dodoi{10.1051/0004-6361:20066744}

\bibitem[{{Voshchinnikov} {et~al.}(2006){Voshchinnikov}, {Il'in}, {Henning}, \& {Dubkova}}]{Voshchinnikov2006}
{Voshchinnikov}, N.~V., {Il'in}, V.~B., {Henning}, T., \& {Dubkova}, D.~N. 2006, \aap, 445, 167, \dodoi{10.1051/0004-6361:20053371}

\bibitem[{{Wakeford} \& {Sing}(2015)}]{WakefordSing2015}
{Wakeford}, H.~R., \& {Sing}, D.~K. 2015, \aap, 573, A122, \dodoi{10.1051/0004-6361/201424207}

\bibitem[{{Welbanks} {et~al.}(2019){Welbanks}, {Madhusudhan}, {Allard}, {Hubeny}, {Spiegelman}, \& {Leininger}}]{Welbanks2019}
{Welbanks}, L., {Madhusudhan}, N., {Allard}, N.~F., {et~al.} 2019, \apjl, 887, L20, \dodoi{10.3847/2041-8213/ab5a89}

\bibitem[{{West} \& {Smith}(1991)}]{West1991}
{West}, R.~A., \& {Smith}, P.~H. 1991, \icarus, 90, 330, \dodoi{10.1016/0019-1035(91)90113-8}

\bibitem[{{Woitke} {et~al.}(2020){Woitke}, {Helling}, \& {Gunn}}]{Woitke2020}
{Woitke}, P., {Helling}, C., \& {Gunn}, O. 2020, \aap, 634, A23, \dodoi{10.1051/0004-6361/201936281}

\bibitem[{{Yair} {et~al.}(1995){Yair}, {Levin}, \& {Tzivion}}]{yair1995}
{Yair}, Y., {Levin}, Z., \& {Tzivion}, S. 1995, \icarus, 114, 278, \dodoi{10.1006/icar.1995.1062}

\bibitem[{{Yu} {et~al.}(2021){Yu}, {He}, {Zhang}, {H{\"o}rst}, {Dymont}, {McGuiggan}, {Moses}, {Lewis}, {Fortney}, {Gao}, {Kempton}, {Moran}, {Morley}, {Powell}, {Valenti}, \& {Vuitton}}]{yu2021}
{Yu}, X., {He}, C., {Zhang}, X., {et~al.} 2021, Nature Astronomy, 5, 822, \dodoi{10.1038/s41550-021-01375-3}

\bibitem[{{Zahnle} {et~al.}(2009){Zahnle}, {Marley}, {Freedman}, {Lodders}, \& {Fortney}}]{Zahnle2009}
{Zahnle}, K., {Marley}, M.~S., {Freedman}, R.~S., {Lodders}, K., \& {Fortney}, J.~J. 2009, \apjl, 701, L20, \dodoi{10.1088/0004-637X/701/1/L20}

\bibitem[{{Zhang}(2020)}]{Zhang2020}
{Zhang}, X. 2020, Research in Astronomy and Astrophysics, 20, 099, \dodoi{10.1088/1674-4527/20/7/99}

\bibitem[{{Zhang} {et~al.}(2013){Zhang}, {West}, {Banfield}, \& {Yung}}]{zhang2013}
{Zhang}, X., {West}, R.~A., {Banfield}, D., \& {Yung}, Y.~L. 2013, \icarus, 226, 159, \dodoi{10.1016/j.icarus.2013.05.020}

\end{thebibliography}
\bibliographystyle{aasjournal}



\end{document}